\documentclass[conf ]{new-aiaa}
\pdfoutput=1
\usepackage[utf8]{inputenc}
\usepackage{textcomp}
\usepackage{placeins}
\usepackage{subcaption}
\usepackage{booktabs}
\usepackage{graphicx}
\usepackage{amsmath}
\usepackage{multirow}
\usepackage[version=4]{mhchem}
\usepackage{siunitx}
\usepackage{longtable,tabularx}
\usepackage[acronym,nonumberlist, nogroupskip]{glossaries}
\usepackage{hyperref}
\usepackage{pdflscape}
\usepackage{rotating}
\usepackage{geometry}
\usepackage{booktabs} 
\usepackage{colortbl}

\hypersetup{
    colorlinks,
    citecolor=black,
    filecolor=black,
    linkcolor=black,
    urlcolor=black
}

\title{Fast Surrogate Models for Adaptive Aircraft Trajectory Prediction in En route Airspace}

\author{Nick Pepper \footnote{Senior Research Associate, The Alan Turing Institute, London, England NW1 2DB, United Kingdom}} 
\affil{The Alan Turing Institute, London, England NW1 2DB, United Kingdom}

\author{Marc Thomas \footnote{Researcher, Department of Research and Development, NATS; also Visiting Professor, Queen Mary University London}}
\affil{NATS, Fareham, England PO15 7FL, United Kingdom}

\author{Zack Xuereb Conti \footnote{Research Fellow, The Alan Turing Institute, London, England NW1 2DB, United Kingdom}}
\affil{The Alan Turing Institute, London, England NW1 2DB, United Kingdom}

\newglossaryentry{atc}{
    name={ATC},
    description={Air traffic control}
}
\newglossaryentry{atco}{
    name={ATCO},
    description={Air traffic control officer}
}
\newglossaryentry{bada}{
    name={BADA},
    description={Base of Aircraft Data}
}
\newglossaryentry{bic}{
    name={BIC},
    description={Bayesian information criterion}
}
\newglossaryentry{bnn}{
    name={BNN},
    description={Bayesian neural network}
}
\newglossaryentry{cas}{
    name={CAS},
    description={Calibrated airspeed. Equivalent to true airspeed when flying at sea level under International Standard Atmosphere conditions}
}
\newglossaryentry{cdf}{
    name={CDF},
    description={Cumulative distribution function}
}
\newglossaryentry{cdo}{
    name={CDO},
    description={Continuous descent operations}
}
\newglossaryentry{esf}{
    name={ESF},
    description={Energy share factor}
}
\newglossaryentry{fl}{
    name={FL},
    description={Flight level}
}
\newglossaryentry{fir}{
    name={FIR},
    description={Flight information region}
}
\newglossaryentry{fpca}{
    name={fPCA},
    description={Functional principal component analysis}
}
\newglossaryentry{gcs}{
    name={GCS},
    description={Geographical coordinate space}
}
\newglossaryentry{ga}{
    name={GA},
    description={General aviation}
}
\newglossaryentry{gan}{
    name={GAN},
    description={Generative Adversarial Network}
}
\newglossaryentry{gmm}{
    name={GMM},
    description={Gaussian mixture model}
}
\newglossaryentry{hmm}{
    name={HMM},
    description={Hidden Markov model}
}
\newglossaryentry{ias}{
    name={IAS},
    description={Indicated airspeed. The \gls{cas} measured by an aircraft's sensors, subject to measurement and position errors}
}
\newglossaryentry{icao}{
    name={ICAO},
    description={International Civil Aviation Organization}
}
\newglossaryentry{imm}{
    name={IMM},
    description={Interacting multiple model}
}
\newglossaryentry{isa}{
    name={ISA},
    description={International standard atmosphere}
}
\newglossaryentry{kf-tp}{
    name={KF-TP},
    description={Kalman filter-based benchmark for the purpose of trajectory prediction, rather than state estimation}
}
\newglossaryentry{ks}{
    name={KS},
    description={Kolmogorov-Smirnov (distance)}
}
\newglossaryentry{kl}{
    name={KL},
    description={Kullback-Leibler (divergence)}
}
\newglossaryentry{lssm}{
    name={LSSM},
    description={Linear state space model}
}
\newglossaryentry{lstm}{
    name={LSTM},
    description={Long short-term memory}
}
\newglossaryentry{lti}{
    name={LTI},
    description={Linear time invariant}
}
\newglossaryentry{lwpf}{
    name={LWPF},
    description={Liu and West particle filter}
}
\newglossaryentry{mad}{
    name={MAD},
    description={Mean absolute deviation}
}
\newglossaryentry{mae}{
    name={MAE},
    description={Mean absolute error}
}
\newglossaryentry{map}{
    name={MAP},
    description={Maximum a posteriori}
}
\newglossaryentry{mse}{
    name={MSE},
    description={Mean squared error}
}
\newglossaryentry{ml}{
    name={ML},
    description={Machine learning}
}
\newglossaryentry{nf}{
    name={NF},
    description={Normalizing flow}
}
\newglossaryentry{rocd}{
    name={ROCD},
    description={Rate of climb/descent}
}
\newglossaryentry{tas}{
    name={TAS},
    description={True airspeed. Aircraft speed relative to the airmass through which it is flying}
}
\newglossaryentry{pde}{
    name={PDE},
    description={Partial Differential Equation}
}
\newglossaryentry{pdf}{
    name={PDF},
    description={Probability density function}
}
\newglossaryentry{pp}{
    name={PP},
    description={Power predictive}
}
\newglossaryentry{smc}{
    name={SMC},
    description={Sequential Monte Carlo}
}
\newglossaryentry{tp}{
    name={TP},
    description={Trajectory Prediction}
}

\makeglossaries

\begin{document}

\maketitle

\begin{abstract}
Trajectory prediction (\gls{tp}) is crucial for ensuring safety and efficiency in  modern air traffic management systems. It is, for example, a core component of conflict detection and resolution tools, arrival sequencing algorithms, capacity planning, as well as several future concepts. However, \gls{tp} accuracy within operational systems is hampered by a range of epistemic uncertainties such as the mass and performance settings of aircraft and the effect of meteorological conditions on aircraft performance. It can also require considerable computational resources.

This paper proposes a method for adaptive \gls{tp} that has two components: first, a fast surrogate \gls{tp} model based on linear state space models (\gls{lssm})s with an execution time that was 6.7 times lower on average than an implementation of the Base of Aircraft Data (\gls{bada}) in Python. It is demonstrated that such models can effectively emulate the \gls{bada} aircraft performance model, which is based on the numerical solution of a partial differential equation (\gls{pde}), and that the \gls{lssm}s can be fitted to trajectories in a dataset of historic flight data. Secondly, the paper proposes an algorithm to assimilate radar observations using particle filtering to adaptively refine \gls{tp} accuracy. Comparison with baselines using \gls{bada} and Kalman filtering demonstrate that the proposed framework improves system identification and state estimation for both climb and descent phases, with 46.3\% and 64.7\% better estimates for time to top of climb and bottom of descent compared to the best performing benchmark model. In particular, the particle filtering approach provides the flexibility to capture non-linear performance effects including the \gls{cas}-Mach transition.

\end{abstract}
\section*{Nomenclature}
\subsection{BADA parameters}
{\renewcommand\arraystretch{1.0}
\noindent\begin{longtable*}{@{}l @{\quad=\quad} l@{}}
$h$  & Geodetic altitude \\
$\frac{dh}{dt}$  & Rate of climb/descent \\
$T$ & ISA temperature \\
$\Delta T$ & Temperature correction \\
$D$ & Aircraft drag\\
$V_{TAS}$ & True airspeed \\
$m$ & Aircraft mass \\
$g_0$ & Gravitational acceleration \\
$f$ & Energy share factor \\
$M$ & Mach number \\
\end{longtable*}}
 \newpage

\subsection{Linear state space models}
{\renewcommand\arraystretch{1.0}
\noindent\begin{longtable*}{@{}l @{\quad=\quad} l@{}}
$A$  & Continuous time dynamics matrix  \\
$B$  & Continuous time input matrix  \\
$\Phi_A$  & Discrete time dynamics matrix  \\
$\Phi_B$  & Discrete time input matrix  \\
$\boldsymbol{x}$ & State vector \\
$\boldsymbol{u}$ & Input vector \\
$L$ & Matrix of scaling factors \\
$\boldsymbol{\theta}$ & LSSM parameters\\
\end{longtable*}}

\subsection{Particle filter parameters}
{\renewcommand\arraystretch{1.0}
\noindent\begin{longtable*}{@{}l @{\quad=\quad} l@{}}
$w$ & Particle weight \\
$n_{eff}$ & Effective particle number \\
$a$ & Shrinkage parameter \\
\end{longtable*}}

\subsection{Kalman filter parameters}
{\renewcommand\arraystretch{1.0}
\noindent\begin{longtable*}{@{}l @{\quad=\quad} l@{}}
$P$ & State uncertainty matrix \\
$Q$ & Process noise matrix \\
$R$ & Measurement noise covariance \\
$\alpha_p, \alpha_q, \alpha_b$ & Scaling terms \\
\end{longtable*}}

\section{Introduction}
\lettrine{T}rajectory prediction (\gls{tp}) is utilised in operational air traffic management systems to underpin traffic load prediction, short and medium term conflict detection, and many other functions. Models such as the Base of Aircraft Data (\gls{bada}) model are deterministic trajectory prediction models calibrated to produce generic aircraft performance \cite{nuic2010user}. Accurate trajectory prediction with deterministic methods is hampered by the presence of significant epistemic uncertainties. For example the mass and performance settings of aircraft are not known to the air traffic control officer (\gls{atco}) \cite{bastas2020data}, while meteorological conditions are uncertain and can have a significant effect on aircraft performance \cite{Bayes+weather}. These epistemic uncertainties can cause significant misspecification between predicted and observed trajectories. 

In order to address this issue, many works have explored leveraging large datasets of available trajectory data in order to improve \gls{tp} using data-driven methods to approach \gls{tp} as a sequence-to-sequence learning task. Investigated methods include Long-Short Term Memory networks \cite{silvestre, N-Incept}, Convolutional Neural Networks \cite{xiang_and_chen}, Generative Adversarial Networks \cite{WU2022103554}, and Hidden Markov Models \cite{pred_analytics}. Such methods tend to be applied to terminal airspace, where there is less diversity in followed routes compared to en route airspace. As an alternative to data-driven \gls{tp} using deterministic data-driven methods, probabilistic methods have been proposed such that predicted trajectories can be obtained through sampling and in principle, credible intervals could be computed to indicate the level of uncertainty in the prediction \cite{gmm, Pepper_DataBADA, hodgkin2025probabilisticsimulationaircraftdescent}. This is useful in the setting of conflict detection, where \gls{tp} methods are often used to identify regions of airspace an aircraft may plausibly occupy within a time interval, which is based on applying conservative bounds around a deterministic prediction to account for model misspecification (see, e.g. \cite{Anderson_Lin_1996, hsu1981evaluation, liu2011}). 

As an alternative to improving the fidelity of the \gls{tp} model, there have been a number of works that seek to assimilate live trajectory data to improve an existing \gls{tp} model. For instance Lymperopoulos \cite{lymperopoulos2010sequential} implemented a sequential conditional particle filter to update trajectory predictions of aircraft in cruise subject to uncertainty arising from wind conditions. Similarly, algorithms have been proposed to estimate aircraft takeoff mass, which is unknown to the \gls{atco} and has a significant effect on climb performance, from trajectory data (see, e.g. \cite{WANG2025109918, sun2016modeling, schultz}). These methods have the advantage of continuously refining predictions based on observations of a trajectory. However, a drawback is that sequential Monte Carlo methods require numerous evaluations of a deterministic \gls{tp} model such as \gls{bada}, which can require considerable computational resources when scaled to hundreds of aircraft within a real-world airspace, even though an individual model evaluation may only be of the order of milliseconds. Instead, the existing \gls{tp} model could be substituted with a fast surrogate model. This would allow rapid sampling for Bayesian assimilation of observed trajectory data, while also enabling probabilistic \gls{tp} by generating trajectory samples inexpensively. 

While Kalman filters are widely used for aircraft tracking (see, e.g. \cite{madyastha2011extended}) they are not necessarily appropriate for the task of system identification of a \gls{tp} model. {Identifying the underlying aircraft dynamics is more challenging than estimating an aircraft's position because of switching between various modes of climb and descent within the same trajectory. Examples of discontinuous changes in aircraft dynamics include the \gls{cas}-Mach transition or when aircraft enter a region of turbulence}. Aircraft performance in descent is often influenced by level-by constraints in the clearances issued by ATCOs. This means that the dynamical model best describing an aircraft's trajectory can change discontinuously between observations. A single Kalman filter is often unable to track systems with such mode switching behaviour \cite{zhang2015}.  For this reason more flexible approaches employing particle filtering \cite{lymperopoulos2008adaptive} or mixtures of linear systems \cite{liu2011} and \gls{tp} models \cite{maeder} have previously been investigated. 

In the proposed method fast sampling is achieved by modelling variations of true airspeed (\gls{tas}) and geodetic altitude in climb as a discrete-time linear (time-invariant) state space model (\gls{lssm}), while a Sequential Monte Carlo (\gls{smc}) method is used as a flexible means of assimilating observations of aircraft trajectories in real time. More specifically, a Liu and West particle filter is implemented for the assimilation as a specialized \gls{smc} algorithm for joint state estimation and system identification \cite{liu2001combined}. A sampling based approximation of the prior distribution for the uncertain aircraft dynamics is formed by fitting \gls{lssm}s to a training dataset of trajectory data in an optimization process. The proposed methodology for adaptive \gls{tp}, which has been developed specifically for online trajectory prediction within a Digital Twin of enroute airspace \cite{DTpaper}, offers several features: 

\begin{itemize}
    \item A fast surrogate for \gls{bada} trajectories using \gls{lssm}s fitted in an optimization process 
    \item A framework for assimilation of trajectory data based on particle filtering that has the flexibility to handle non-linear changes in aircraft dynamics, as occurs at the \gls{cas}-Mach transition for instance
    \item  An improved mean prediction of the top of climb point through assimilation of trajectory data 
    \item An ensemble \gls{tp} through sampling of particles in the filter that may be used to compute credible intervals for the \gls{tp}
\end{itemize}

The remainder of this paper follows the following structure: Section~\ref{sec:bada_ssm} describes the proposed fast \gls{lssm} surrogate as a substitute for \gls{bada} and optimization process for fitting trajectory data. Section~\ref{sec:particle_filter} outlines the steps of the particle filter approach used to assimilate trajectory observations as they become available and update an ensemble trajectory prediction. Section~\ref{sec:data_prep} outlines the datasets used to fit \gls{lssm}s to form a prior on the system dynamics and the held out dataset used to test the effectiveness algorithm. Finally, section~\ref{sec:results} presents the results of applying the proposed algorithm to the held out dataset, baselines by \gls{bada} and some standard data-driven approaches.


\section{Methodology}
\subsection{State-space modelling of the BADA equations}\label{sec:bada_ssm}
The Base of Aircraft Data (\gls{bada}) is a total-energy model that uses a physics-based model to predict aircraft trajectories for a range of aircraft types. The model is based around a \gls{pde} that balances work done by the aircraft thrust against changes in gravitational potential energy and kinetic energy. By assuming either a constant calibrated airspeed (\gls{cas}) or constant Mach speed with altitude, an equation for rate of climb or descent (\gls{rocd}) can be defined \cite{nuic2010user}:
\begin{equation}
    \frac{dh}{dt} = \frac{T-\Delta T}{T} \Big[ \frac{(T_{HR}-D)V_{TAS}}{mg_0} \Big] f(M), \label{eq:bada_rocd}
\end{equation}
where $h$ denotes altitude, $\frac{dh}{dt}$ denotes the \gls{rocd}, $T_{HR}$ the aircraft thrust, $V_{TAS}$ the TAS, and $f(\cdot)$ the energy share factor (\gls{esf}), defined as a function of the Mach number, $M$. The \gls{esf} governs the tradeoff between changes in gravitational potential energy and kinetic energy. It has a different functional form depending on whether the aircraft is flying at constant \gls{cas} or Mach speed. Additionally, $T$ represents air temperature in the International Standard Atmosphere (\gls{isa}) model and $\Delta T$ a temperature correction that may be applied to \gls{bada}, although in what follows this is set to 0.

Many of the terms in \eqref{eq:bada_rocd} are functions of $h$ and $V_{TAS}$. For instance, $T$ varies with $h$ in the \gls{isa} model while in \gls{bada} $T_{HR}$ is a function of $h$ for jet aircraft and both $h$ and $V_{TAS}$ for turboprop aircraft. Similarly, $D$ is a function of $V_{TAS}^2$ but also has further dependence on $h$ and $V_{TAS}$ through the aerodynamic coefficients. Lastly, $f(\cdot)$ is a four branch function where each branch is a function of $M$, which itself can be defined as a conversion from $V_{TAS}$ at a specific $h$. 

We define a linear time-invariant (\gls{lti}) model for climbing aircraft, with a two-dimensional state vector including the geodetic altitude, and \gls{tas}, denoted $\boldsymbol{x}=[h, V_{TAS}]^\top$. These quantities are selected as the minimum set of variables required by \gls{bada}. The continuous time state-space equation for this system is defined as:
\begin{equation}
    \dot{\boldsymbol{x}}=A\boldsymbol{x}+B\boldsymbol{u},
    \label{eq:ssm0}
\end{equation}
where $A\in\Re^{2\times 2}$ is the dynamics matrix, $B\in\Re^{2 \times 2}$ is the input matrix, and $u\in\Re^{2}$ is the input (or control) vector. \eqref{eq:ssm0} is a continuous-time model. However, radar measurements are available at discrete times, for instance Mode S radar returns refresh every 6s. For this reason the discrete time formulation of \eqref{eq:ssm0} is used:
\begin{equation}
    {\boldsymbol{x}}^{(t+1)}=\Phi_A\boldsymbol{x}^{(t)}+\Phi_B\boldsymbol{u}_t,
    \label{eq:ssm}
\end{equation}
where $\Phi_A$ and $\Phi_B$, the discrete time versions of $A$ and $B$ respectively, are computed from $A$, assuming it is invertible, via:
\begin{equation}
    \Phi_A=\text{e}^{\Delta tA} \;\text{and}\; \Phi_B=(\Phi_A-I_2)A^{-1}B,
    \label{eq:disc_time}
\end{equation}
where $I_2$ is a $2\times 2$ identity matrix and radar blips are 6 seconds apart i.e. $\Delta t=6$s. Following the developments in Appendix~\ref{sec:app_A}, it was found that the forcing term in the linearization of \eqref{eq:bada_rocd} was dependent on the state, rather than time. Rather than remove the forcing term entirely we retain a constant forcing operator, with \eqref{eq:ssm} recast as:

\begin{equation}
    {\boldsymbol{x}}^{(t+1)}=\Phi_A\boldsymbol{x}^{(t)}+\Phi_B. 
    \label{eq:ssm2}
\end{equation}
The \gls{lssm} described by \eqref{eq:ssm2} can be interpreted as having a forcing applied from the aircraft thrust that increases both \gls{tas} and altitude (when flying at constant \gls{cas}), while the state space operator $\Phi_A$ describes the dynamics that govern the tradeoff between \gls{tas} and $h$, in a similar manner to the \gls{esf} in \eqref{eq:bada_rocd}. 

System identification was performed through an optimization procedure that determined the elements of $\Phi_A$ and $\Phi_B$ from trajectory data. Let the set of $n$ radar returns that represent an aircraft trajectory, either obtained from real-world data or from solution of the \gls{bada} equations, be denoted $\mathbb{X}=[\boldsymbol{x}^{(1)},\dots, \boldsymbol{x}^{(n)}]$. The fitted \gls{lssm} model for $\mathbb{X}$ was found through solving the unconstrained optimization problem: 

\begin{equation}
   \hat{\boldsymbol{\theta}}=\underset{\boldsymbol{\theta}}{\text{arg min}}\;\mathcal{J}(\boldsymbol{\theta}|\mathbb{X}),
   \label{eq:opt}
\end{equation}
where $\boldsymbol{\theta}\in\Re^6$ collects the elements of $\Phi_A$ and $\Phi_B$. The cost function, $\mathcal{J}(\cdot)$ is defined as:

\begin{equation}
    \mathcal{J}(\boldsymbol{\theta}|\mathbb{X})=\sum_{i=2}^{n} (\hat{\boldsymbol{x}}^{(i)}-\boldsymbol{x}^{(i)})^\top L^{-2} (\hat{\boldsymbol{x}}^{(i)}-\boldsymbol{x}^{(i)}),
       \label{eq:opt2}
\end{equation}
where $L\in\Re^{2\times2}$ is a diagonal matrix containing scaling factors. In this work 30,000~ft and 400~kts were used as scaling factors, reflecting typical cruising altitude and groundspeed for jet engined aircraft in en route airspace. The set $\hat{\mathbb{X}}=[\boldsymbol{x}^{(1)}, \hat{\boldsymbol{x}}^{(2)}, \dots, \hat{\boldsymbol{x}}^{(n)}]$ is generated by initializing at $\boldsymbol{x}^{(1)}$ (the first radar blip in $\mathbb{X}$) and recursively applying $\Phi_A$ and $\Phi_B$, as in \eqref{eq:ssm}. Given that the derivatives of $\mathcal{J}(\boldsymbol{\theta}|\mathbb{X})$ with respect to $\boldsymbol{\theta}$ are non-trivial {due to the recursive summation in $\hat{\mathbb{X}}$, the Nelder-Mead algorithm, a non-convex optimization method, was used to solve \eqref{eq:opt} \cite{nelder1965simplex}. 

Figure~\ref{fig:bada_approx} shows the result of applying the the \gls{lssm} methodology described here to emulate the trajectories in \gls{bada} of commonly occurring aircraft in UK airspace. The selected aircraft types contain a mix of ICAO wake turbulence categories and engine types. Trajectory data, $\mathbb{X}$, is generated using \gls{bada} for climbs and descents between flight level 210 (21,000~ft at standard atmospheric pressure) and either the geodetic altitude corresponding to the tropopause or 5,000~ft above the \gls{cas}-Mach transition point. In Figure~\ref{fig:bada_approx} the \gls{bada} trajectories are solid lines, while dashed lines indicate the trajectories generated from recursively applying the \gls{lssm}s ($\hat{\mathbb{X}}$). As can be seen from the figure, there is a high degree of agreement between the two set of trajectories. However, the trajectories from the \gls{lssm}s are less expensive to generate as they do not require \eqref{eq:bada_rocd} to be solved numerically.  

Table~\ref{tab:bada_climb} indicates the root mean square error between the trajectories derived from \gls{bada} versus those from the fitted \gls{lssm}s for the aircraft types in Figure~\ref{fig:bada_approx} in climb. Using the initial conditions for the B738 to normalise the scales for altitude and airspeed (210,000 ft and 400 kts), these errors can be expressed as a percentage. It was found that the percentage error of the \gls{lssm} surrogates is very low: 0.021\% error in the geodetic altitude and 0.028\% for \gls{tas}. Similarly, Table~\ref{tab:bada_descent} displays the results of the fitting for BADA trajectories in descent. Comparing the data in Tables~\ref{tab:bada_climb} and ~\ref{tab:bada_descent}, the mean fitting error is generally worse for descent. For trajectory segments above the transition point this is likely because the descending trajectories are longer due to the transition point being lower in descent, hence the descents accumulate more error. Below the transition point, the mean error in descent is raised by the outlier of the PC12 aircraft type. The parameters of the PC12 in \gls{bada} are such that there is a discontinuous change in the thrust profile around \gls{fl} 230. As a consequence, the PC12 trajectory is piecewise linear below the transition point, which cannot be well approximated using a single \gls{lssm}. Lastly, Tables~\ref{tab:bada_climb} and ~\ref{tab:bada_descent} show the relative difference in execution time between the Python implementation of \gls{bada} used and the fitted \gls{lssm}s, with the \gls{lssm}s faster by a factor or 5.26 in climb and 8.11 in descent (descending trajectories were generally longer than climbs).


\begin{figure}
    \centering
    \begin{subfigure}[b]
    {\textwidth}
    \centering
    \includegraphics[width=0.8\linewidth]{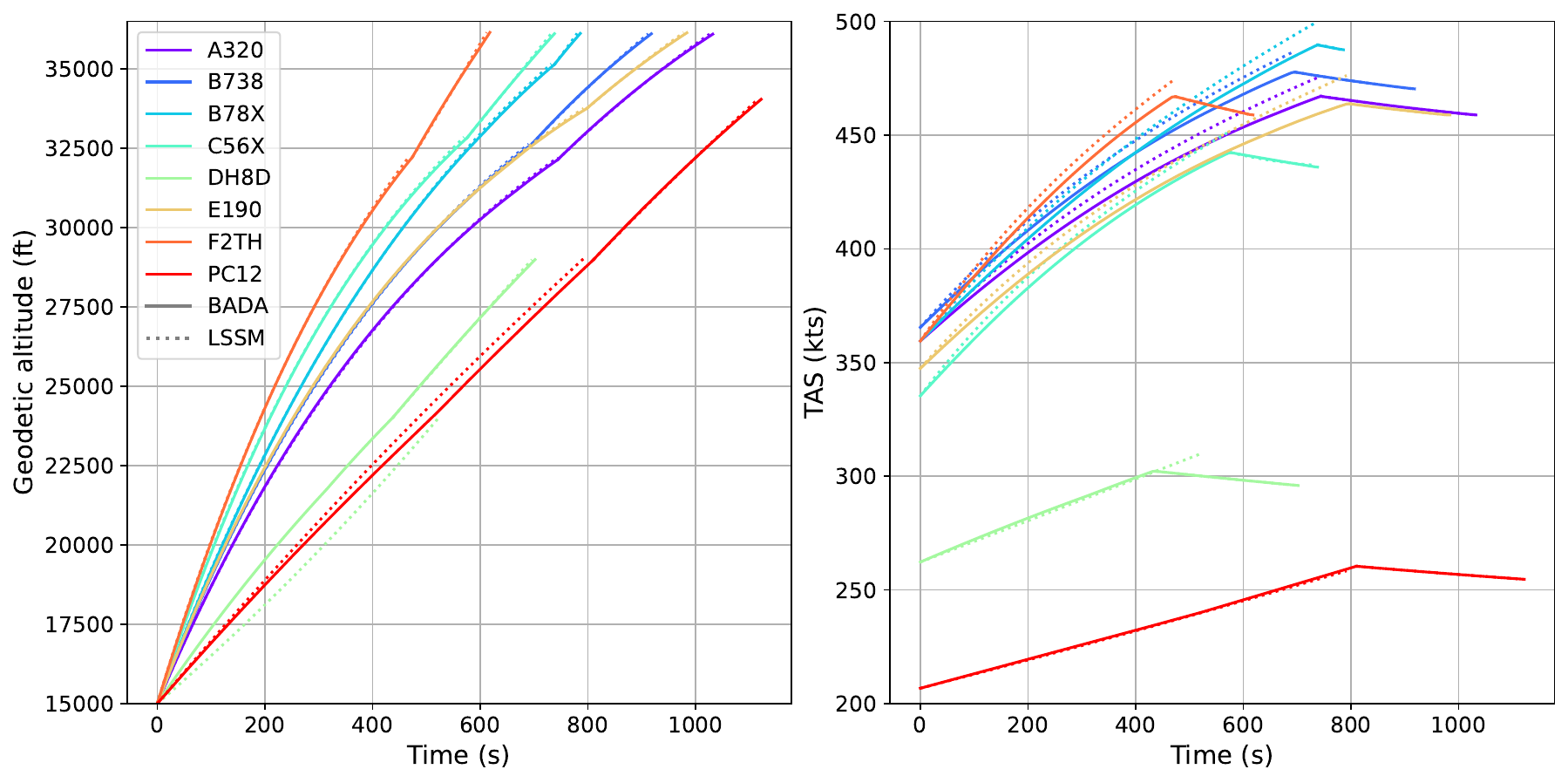}
    \caption{}
    \label{fig:bada_approx_a}
    \end{subfigure}
    \begin{subfigure}[b]
    {\textwidth}
    \centering
    \includegraphics[width=0.8\linewidth]{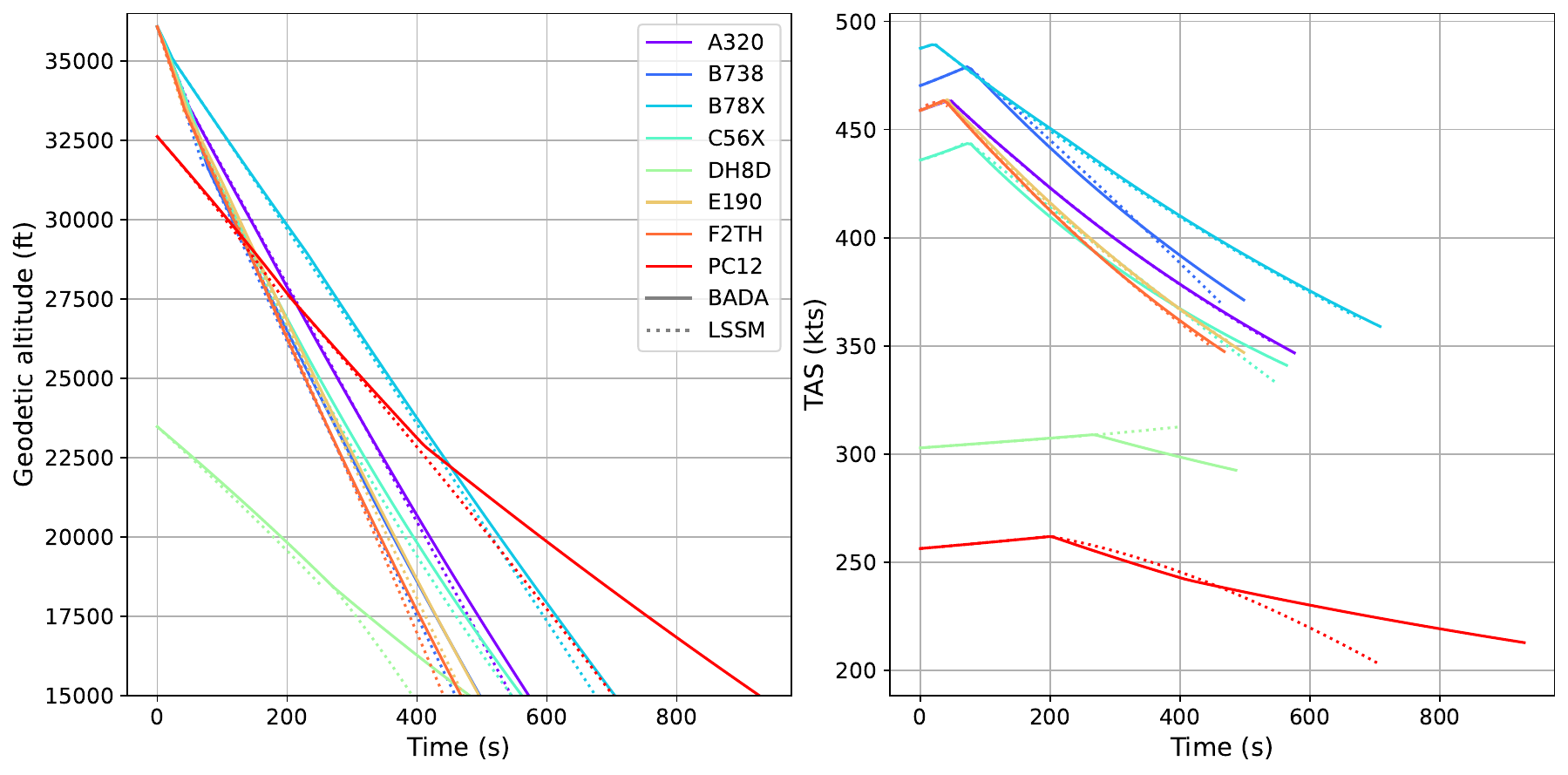}
    \caption{}
    \label{fig:bada_approx_b}
    \end{subfigure}
    \caption{Fitted \gls{lssm}s used as an emulator for the trajectories of a range of climbing and descending aircraft in the \gls{bada} model.}
    \label{fig:bada_approx}
\end{figure}

\begin{table}
\begin{center}
\begin{tabular}{c|cc|cc|ccc}
\toprule
Aircraft type & \multicolumn{2}{c|}{Below \gls{cas}-Mach transition} & \multicolumn{2}{c|}{Above \gls{cas}-Mach transition}  & \multicolumn{3}{c}{Execution time (ms)}\\
& $\Delta h$ (ft) & $\Delta V_{TAS}$ (kts) & $\Delta h$ (ft) & $\Delta V_{TAS}$ (kts) & BADA & LSSM & BADA/LSSM\\
\midrule
A320 & 31.29 & 5.43 & 23.57 & 0.00 & 5.72 & 0.93 & 6.15 \\
B738 & 34.73 & 5.41 & 26.64 & 0.00 & 4.67 & 0.87 & 5.37 \\
B78X & 43.15 & 6.12 & 61.05 & 0.10 & 3.54 & 0.70 & 5.07 \\
C56X & 44.54 & 5.56 & 44.53 & 0.32 & 3.35 & 0.64 & 5.26 \\
DH8D & 1362.43 & 0.88 & 36.46 & 0.00 & 3.17 & 0.71 & 4.44 \\
E190 & 29.94 & 7.15 & 28.95 & 0.00 & 4.58 & 0.85 & 5.40 \\
F2TH & 48.31 & 4.56 & 60.79 & 0.03 & 2.75 & 0.53 & 5.15 \\
PC12 & 302.42 & 0.41 & 30.77 & 0.00 & 4.83 & 0.93 & 5.21 \\
\midrule
Mean & 237.10 & 4.44 & 39.10 & 0.06 & 4.07 & 0.77 & 5.26 \\
\bottomrule
\end{tabular}
\end{center}
\caption{Root mean square error between \gls{bada} trajectories and those generated from the fitted \gls{lssm}s for aircraft in climb.}
\label{tab:bada_climb}
\end{table}


\begin{table}
\begin{center}
\begin{tabular}{c|cc|cc|ccc}
\toprule
Aircraft type & \multicolumn{2}{c|}{Below \gls{cas}-Mach transition} & \multicolumn{2}{c|}{Above \gls{cas}-Mach transition}  & \multicolumn{3}{c}{Execution time (ms)}\\
& $\Delta h$ (ft) & $\Delta V_{TAS}$ (kts) & $\Delta h$ (ft) & $\Delta V_{TAS}$ (kts) & BADA & LSSM & BADA/LSSM\\
\midrule
A320 & 300.09 & 0.23 & 38.04 & 1.13 & 4.20 & 0.57 & 7.36 \\
B738 & 740.26 & 3.38 & 272.22 & 0.02 & 4.87 & 0.62 & 7.89 \\
B78X & 330.64 & 0.80 & 171.34 & 0.01 & 7.53 & 0.88 & 8.59 \\
C56X & 345.38 & 4.24 & 259.19 & 0.32 & 4.65 & 0.65 & 7.16 \\
DH8D & 778.10 & 7.95 & 190.10 & 0.08 & 5.59 & 0.63 & 8.83 \\
E190 & 401.66 & 0.73 & 223.11 & 0.10 & 5.00 & 0.65 & 7.68 \\
F2TH & 375.93 & 0.56 & 32.27 & 1.10 & 3.77 & 0.53 & 7.18 \\
PC12 & 1519.78 & 8.34 & 148.32 & 0.02 & 15.72 & 1.54 & 10.23 \\
\midrule
Mean & 598.98 & 3.28 & 166.82 & 0.35 & 6.42 & 0.76 & 8.11 \\
\bottomrule
\end{tabular}
\end{center}
\caption{Root mean square error between \gls{bada} trajectories and those generated from the fitted \gls{lssm}s for aircraft in descent.}
\label{tab:bada_descent}
\end{table}

\FloatBarrier

\subsection{Particle filtering}\label{sec:particle_filter}
Repeating the optimization procedure in \eqref{eq:opt} for a dataset of $n_t$ trajectories, denoted $\mathbb{T}=[\mathbb{X}^{(1)}, \dots, \mathbb{X}^{(n_t)}]$, yields a set of $n_t$ optimal \gls{lssm}s, which we denote $\mathbb{S}=[(\Phi_A, \Phi_B)^{(1)},\dots (\Phi_A, \Phi_B)^{(n_t)}]$. The LSSMs are optimal in the sense that they are the predictors that most closely match the trajectories in $\mathbb{T}$. The trajectories in $\mathbb{T}$ and LSSMs in $\mathbb{S}$ are specific to an aircraft type. The \gls{lssm}s in $\mathbb{S}$ can be thought of as a set of samples drawn from the prior distribution for $\boldsymbol{\theta}$, $p(\boldsymbol{\theta})$. As more observations of an unseen trajectory become available, a particle filter approach is used to jointly estimate both the state of the aircraft, $p(\boldsymbol{x}|\boldsymbol{y}^{(1)},\dots, \boldsymbol{y}^{(t)})$, and identify the underlying dynamics of the trajectory $p(\boldsymbol{\theta}|\boldsymbol{y}^{(1)},\dots, \boldsymbol{y}^{(t)})$, where $\boldsymbol{y}^{(i)}$ denotes an observation of altitude and \gls{tas} from the $i^{\text{th}}$ radar blip in the trajectory and recalling from the previous sections that $\boldsymbol{\theta}$ collects the elements of $\Phi_A$ and $\Phi_B$. At a given timestep $t$, we wish to estimate the future states of the aircraft and the associated uncertainty. 

A Liu and West particle filter (\gls{lwpf}) \cite{liu2001combined} was implemented as a specialized method for joint state estimation and system identification. The Liu and West filter differs from methods such as the Interacting Multiple Model (\gls{imm}) approach in that it admits continuous system parameters, $\boldsymbol{\theta}$ \cite{mazor1998}. In contrast, approaches such as \gls{imm} filter from a discrete set of \gls{lssm}s in parallel which might limit the ability of the filter to assimilate radar observations if the underlying system parameters are not present in the set of models. The primary steps of the algorithm are summarized below, the interested reader is referred to Liu and West \cite{liu2001combined} and Nemeth et al \cite{nemeth2013sequential} for more details on the method. In what follows, $\boldsymbol{x}^{(0)}$ denotes an initial observation of the aircraft's state which is used to initialise the filter. The prior distribution $p(\boldsymbol{x}^{(0)})$ represents the uncertainty in the initial state given measurement error. As is discussed in the context of Kalman filtering in Appendix~\ref{sec:app_kalman}, this uncertainty is significantly smaller than the uncertainty surrounding the system dynamics. The main steps of the proposed adaptive trajectory prediction algorithm can be summarized as:

\begin{enumerate}\addtocounter{enumi}{-1}
\item \textbf{Sampling}

Draw $n_p$ samples from the prior distribution for $p(\boldsymbol{x}^{(0)})$ and sample from $\mathbb{S}$ with replacement $n_p$ times to obtain a set of particles with a state and \gls{lssm} associated with each. Initialize the particle weights, $\boldsymbol{w}^{(0)} = \boldsymbol{1}/{n_p}$.

\item \textbf{Prediction}

Evolve the state of each particle using the associated \gls{lssm} using \eqref{eq:ssm2}. 

\item \textbf{Shrink system parameters towards the mean}

Compute average system parameter from weighted sum of particles:
\begin{align}
    \bar{\boldsymbol{\theta}} = \sum_{i=1}^{n_p} \boldsymbol{w}_i^{(t-1)} \boldsymbol{\theta}_i^{(t-1)},
\end{align}
together with the variance of system parameters in the filter: 
\begin{align}
    V^{(t-1)} = \sum_{i=1}^{n_p}\boldsymbol{w}^{(t-1)} (\boldsymbol{\theta}_i^{(t-1)}-\bar{\boldsymbol{\theta}})(\boldsymbol{\theta}_i^{(t-1)}-\bar{\boldsymbol{\theta}})^\top,
\end{align}
to yield the updated system parameters using associated with the $i$\textsuperscript{th} particle:

\begin{align}
    \boldsymbol{\theta}_i^{(t)} = a\boldsymbol{\theta}_i^{(t-1)} + (1-a)\bar{\boldsymbol{\theta}}+N(\boldsymbol{0}, b^2V^{(t-1)}),
\end{align}
where $a=1-b^2$, which controls shrinkage towards $\bar{\boldsymbol{\theta}}$. We achieved best results using $b=0.2$.

\item \textbf{Update particle states and weights}

Using the measurement $\boldsymbol{y}^{(t)}$, compute the observation likelihood for each particle through: 
\begin{align}
    p(\boldsymbol{y}^{(t)}| \boldsymbol{x}_i^{(t)}, \boldsymbol{\theta}_i^{(t)})=\text{e}^{-\frac{1}{2}\big(({\boldsymbol{y}}^{(t)}-\boldsymbol{x}_i^{(i)})^\top R^{-1} ({\boldsymbol{y}}^{(t)}-\boldsymbol{x}_i^{(t)})\big)},
\end{align}
where $\boldsymbol{x}_i^{(t)}$ denotes the state of the $i$\textsuperscript{th} particle at timestep $t$ and $\boldsymbol{\theta}_i^{(t)}$ the parameters of the \gls{lssm} associated with that particle. The matrix $R$ is as defined in \eqref{eq:opt2}. The observation likelihood is used to update the normalized particle weight through:  
\begin{align}
    w_i=\frac{p(\boldsymbol{y}^{(t)}| \boldsymbol{x}_i^{(t)}, \boldsymbol{\theta}_i^{(t)})}{\sum_{j=1}^{n_p}p(\boldsymbol{y}^{(t)}| \boldsymbol{x}_j^{(t)}, \boldsymbol{\theta}_j^{(t)})}.
\end{align}
\newpage
\item \textbf{Resampling}

Resample the particle filter, where the probability of selecting the $i$\textsuperscript{th} particle is set to $w_i$. To avoid particle degeneracy, where weight is concentrated in only a few particles, resampling through stratified sampling is performed if the effective number of particles, $n_{eff}$, drops below a threshold. The effective number of particles is defined as: 
\begin{align}
    n_{eff} = \frac{1}{\sum_{i=1}^{n_p}w_i^2},
\end{align}
with the threshold below which stratified resampling is performed set to $0.5\times n_p$. Occasionally aircraft will change their mode of climb or descent, leading to rapid changes in performance. This necessitated re-initialising the particle filter. This was triggered if the absolute difference between the estimated \gls{tas} and last observed \gls{tas} was greater than 5 knots.

\item \textbf{State estimation}

The filter state is the weighted sum of the states of individual particles, i.e.:
\begin{align}
    \hat{\boldsymbol{x}}^{(t)} = \sum_{i=1}^{n_p}w_i\boldsymbol{x}^{(t)}_i.
\end{align}
The estimated state of the filter is used to determine whether to perform ensemble trajectory prediction.  

\item \textbf{Ensemble Trajectory Prediction}

If the estimated particle altitude, $\hat{\boldsymbol{x}}^{(t)}_1$, is equal to or greater than the aircraft's target altitude then cease iteration. Otherwise, a prediction for future states of the aircraft's climb is made using multinomial sampling from the particles.  Initializing at $\boldsymbol{x}^{(t)}_i$, the \gls{lssm} associated with that particle is used to generate a predicted trajectory, $[\mathbf{x}^{(t)}_i, \mathbf{x}^{(t+1)}_i, \dots, \mathbf{x}^{(t+n)}_i]$. The mean and standard deviations of these samples can be used to return a deterministic \gls{tp} together with a credible interval. The algorithm then returns to step 1. 

\end{enumerate}

\FloatBarrier
\section{Data preparation}\label{sec:data_prep}
A dataset consisting of 29 days' worth of Mode S radar surveillance data, collected from en route airspace in the UK in September 2019, was used to assess the effectiveness of the adaptive \gls{tp} algorithm. The dataset was split into training, validation, and test sets in a 70/10/20 ratio. This resulted in 21 days of training data, 2 days of validation data, and 6 days of test data. The dataset was split based in this way to minimise data leakage between the train and test sets, particularly with regard to meteorological conditions.  

The optimization process described in Section~\ref{sec:bada_ssm} was used to generate the set of \gls{lssm}s, $\mathbb{S}$, for the days in the training dataset. Table~\ref{tab:datasets} displays the investigated aircraft types and the number of trajectories for each aircraft type in the various datasets. A range of aircraft types were investigated, from large passenger jets such as the B738 and A320 which are commonplace in UK airspace, to smaller business jets such as the C56X, and turboprop aircraft which are less frequently occurring and have markedly different performance characteristics to jet aircraft. 

The performance of the particle filtering-based method for state estimation and trajectory prediction introduced in Section~\ref{sec:particle_filter} was benchmarked against a simple method based on Kalman filtering. This simplistic trajectory prediction method assumes the aircraft will continue to climb or descend at the filtered ROCD and TAS and is outlined in Appendix~\ref{sec:app_kalman}. The validation dataset was used to perform a sweep to set the hyperparameters of this method. Similarly, the number of particles used in the particle filter was selected based on a sweep of possible particle numbers. The results of this sweep are presented in Appendix~\ref{sec:app_lwpf}. Based on these results a maximum of 400 particles were used for each aircraft type (maximum because some aircraft types had fewer than 400 particles in their training datasets).

\begin{table}
\centering
\begin{tabular}{c|c|cc|cc|cc|cc}
\toprule
Aircraft type & Engine type & \multicolumn{2}{c|}{Training trajectories} & \multicolumn{2}{c|}{Validation trajectories} & \multicolumn{2}{c|}{Test trajectories} & \multicolumn{2}{c}{Total} \\ 
 && Climb & Descent & Climb & Descent & Climb & Descent & Climb & Descent\\
\midrule
B738 & Jet & 16931 & 16476 & 1627 & 1622 & 4710 & 4526 & 23268 & 22624 \\
A320 & Jet & 11198 & 12282 & 1050 & 1204 & 3169 & 3547 & 15417 & 17033 \\
E190 & Jet & 2409 & 2443 & 160 & 174 & 715 & 701 & 3284 & 3318 \\
C56X & Jet & 342 & 372 & 36 & 40 & 108 & 101 & 486 & 513 \\
B78X & Jet & 61 & 61 & 5 & 6 & 15 & 12 & 81 & 79 \\
DH8D & Turboprop & 2660 & 1878 & 254 & 153 & 759 & 543 & 3673 & 2574 \\
F2TH & Turboprop & 162 & 216 & 18 & 16 & 53 & 58 & 233 & 290 \\
PC12 & Turboprop & 121 & 94 & 19 & 19 & 37 & 32 & 177 & 145 \\
\bottomrule
\end{tabular}
\caption{Number of trajectories in the train, validation, and test datasets for each aircraft type.}
\label{tab:datasets}
\end{table}

\section{Results}\label{sec:results}
The adaptive \gls{tp} algorithm presented in Section~\ref{sec:particle_filter} was run over the trajectories in the test dataset for each aircraft type in Table~\ref{tab:datasets}. The accuracy of the proposed method was assessed by predicting the top of climb/bottom of descent point, in terms of the time taken to achieve that altitude and the distance flown while doing so. An ensemble of trajectories was generated by the \gls{lwpf}, with the mean prediction of top of climb/bottom of descent of the samples compared against the test data. Three methods were used to benchmark the adaptive \gls{tp} algorithm, two of which based around \gls{bada}, while the third was a simple benchmark employing Kalman filtering which is further described in Appendix~\ref{sec:app_kalman}. To emphasize that the purpose of the filter in this benchmark is for \gls{tp}, \emph{not} state estimation, in what follows \gls{kf-tp} refers to the Kalman filter-based benchmark, while \gls{lwpf} refers to the proposed algorithm in Section~\ref{sec:particle_filter}. 

Further to the \gls{kf-tp} benchmark, two different methods of generating trajectories using \gls{bada} were included as benchmark \gls{tp} methods: one in which the \gls{bada} trajectory was generated at the first timestep and `frozen' for the remainder of the trajectory and a second implementation where \gls{bada} was re-initialized at the measured flight level at each timestep. The \gls{bada} trajectory at $t=0$ was used to quantify the level of misspecification between existing deterministic \gls{tp} methods and real-world trajectories. Re-initializing \gls{bada} every timestep provides a benchmark adaptive \gls{tp} method that adjusts the initial altitude of the aircraft using the last observation, but does not assimilate any aspect of aircraft performance. 


\begin{table}
\centering
\begin{tabular}{l|cc|cc}
\toprule
Method & \multicolumn{2}{c|}{MAE time error (s)} & \multicolumn{2}{c}{MAE distance error (nmi)} \\
& Climb & Descent & Climb & Descent \\
\midrule
BADA ($t=0$) & 95.61 & 112.15 & 59.66 & 19.48 \\
BADA & 12.70 & 9.88 & 43.04 & 12.70 \\
KF-TP & 9.66 & 18.59 & 7.70 & 14.26  \\
LWPF & \cellcolor[RGB]{179,255,179}5.19 & \cellcolor[RGB]{179,255,179}6.56 & \cellcolor[RGB]{179,255,179}3.94 & \cellcolor[RGB]{179,255,179}4.27 \\
\bottomrule
\end{tabular}
\caption{\gls{mae} to time and location of climb/bottom of descent for the investigated \gls{tp} methods.}
\label{tab:results_by_method}
\end{table}

\begin{figure}
    \centering
     \includegraphics[width=1.0\linewidth]{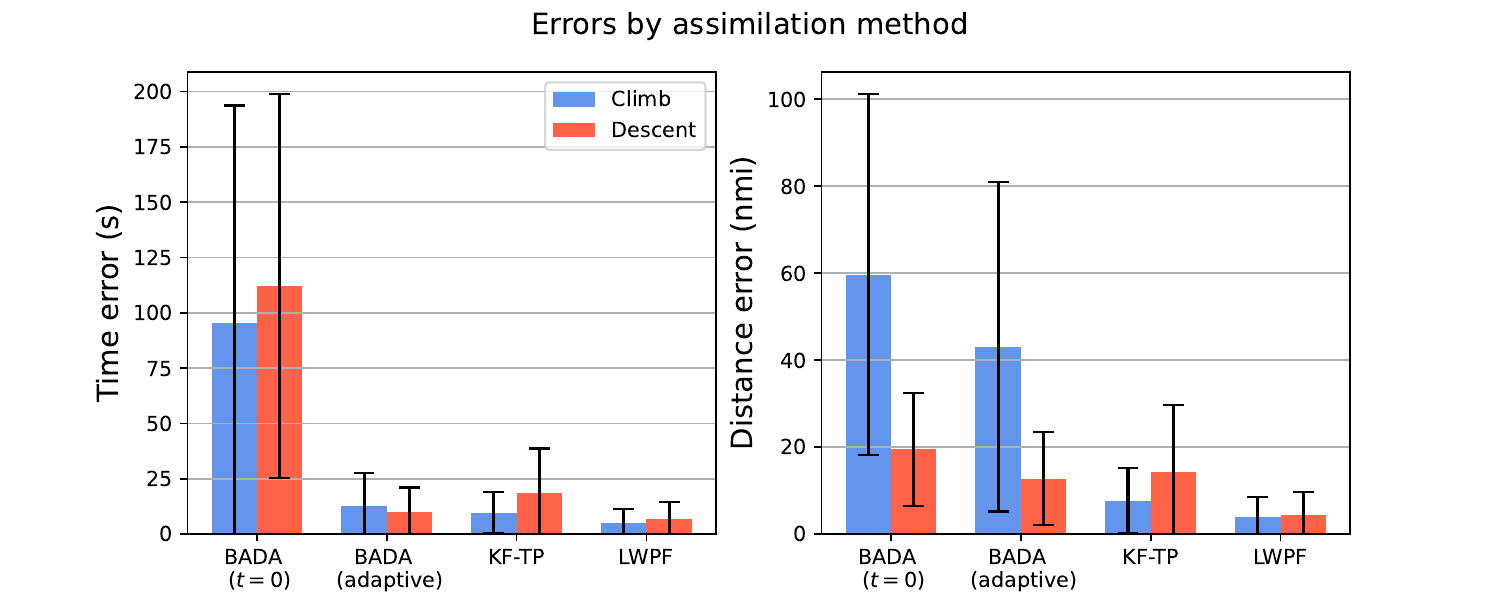}   
    \caption{Comparison of the \gls{mae} in predicting the time and location of top of climb/bottom of descent between the investigated \gls{tp} methods.}
    \label{fig:error_bars_by_method}
\end{figure}

Figure~\ref{fig:error_bars_by_method} compares the performance of the various \gls{tp} methods, as quantified by the mean absolute error (\gls{mae}). The results are tabulated in Table~\ref{tab:results_by_method}. As expected, the three adaptive \gls{tp} methods greatly outperform the \gls{bada} trajectory at $t=0$. Due to local constraints and procedures, as well as other aleatoric and epistemic uncertainties, there is significant mis-specification of aircraft speed profiles in \gls{bada}, leading to large errors in the flown distance, particularly in climb. The error bars for the various \gls{bada} based methods are significantly wider than for the \gls{kf-tp} and \gls{lwpf} approaches suggesting that \gls{bada} does not have the flexibility to account for the range of aircraft performance observed in the test dataset. The \gls{kf-tp} and \gls{lwpf} based methods are more accurate than the two benchmarks using \gls{bada}. Predictions of descent are generally less accurate in the \gls{kf-tp} and \gls{lwpf} models compared to climb, which is understandable given that descents are heavily conditioned on procedural constraints, which are not explicitly modeled here.

From Figure~\ref{fig:error_bars_by_method} and Table~\ref{tab:results_by_method} it can be seen that the \gls{lwpf} has the lowest \gls{mae} for both the estimated time and flown distance of the investigated methods. We define the flown distance as the integrated \gls{tas} over the trajectory, to correct for wind effects. The \gls{lwpf} approach has time estimation errors 46.3\% and 64.7\% lower, and distance errors 48.8\% and 70.1\% lower for climb and descend than the baseline \gls{kf-tp} method. To gain more insight into this result the \gls{mae} for the \gls{kf-tp} approach is compared against that of the \gls{lwpf} method by aircraft type in Figure~\ref{fig:kf_lwpf_by_actype} and Table~\ref{tab:kf_versus_lwpf}. Figure~\ref{fig:kf_versus_lwpf_test_size} provides another visualisation of the data, with the marker size indicating the number of radar blips in the test dataset (recall that the models were used to predict top of climb/bottom of descent after every new observation). Points in the top right indicate that the \gls{lwpf} \gls{mae} is lower than that of the \gls{kf-tp}, while the bottom left quadrant indicates the \gls{mae} is lower for the \gls{kf-tp} for both time and distance flown. From the figures and table it is clear that the \gls{mae} for jet aircraft is generally lower than that for turboprops (DH8D, F2TH, and PC12). The percentage difference between the \gls{lwpf} and \gls{kf-tp} is greatest for the largest passenger jet types such as the A320, B738, and E190, suggesting that the performance of the particle filter is strongly correlated with the diversity of the particles in its prior, that is the representation of that aircraft type in the data. This observation would seem to be supported by the results for the B78X, which is expected to have similar performance characteristics to the A320 and B738, but with significantly fewer aircraft in the training dataset (see Table~\ref{tab:datasets}). Comparing the B78X errors to the B738, the B78X error is consistently greater for the \gls{lwpf} in Table~\ref{tab:kf_versus_lwpf}, whereas the \gls{kf-tp} error for the B78X is either comparable to, or in several cases better than, the B738. Prediction of the DH8D and PC12 turboprops in descent is more accurate for the \gls{kf-tp}, but not for the F2TH. The F2TH and PC12 occur roughly as frequently as one another, suggesting that the reasons for the relatively poor performance of the \gls{lwpf} is not necessarily explained by insufficient numbers of particles in the training set. Turboprop aircraft frequently descend at a fixed rate of descent, which can be better approximated by the fixed \gls{rocd} trajectory generation model of the \gls{kf-tp} \cite{hodgkin2025probabilisticsimulationaircraftdescent}. 

Finally, Figure~\ref{fig:ens_pred_climb} plots the performance of the methods for a set of climbing flights in the test dataset. Similarly, Figure~\ref{fig:ens_pred_descent} displays trajectories for these aircraft types in descent. One flight from the A320, C56X, and PC12 datasets was chosen for visualisation to give examples of aircraft with markedly different (physical) performance characteristics. The A320 is a large passenger jet that frequently occurs in UK airspace, the C56X is a smaller business jet, and the PC12 an example of a single engined turboprop aircraft that is the least common of the investigated aircraft types. Each sub-figure is split into left and right panels, with the left panels plotting aircraft altitude against time and the right panel displaying \gls{tas} against time. In the left panels, the mean predicted trajectory from the particles in the \gls{lwpf} are plotted every one or two minutes and compared to the generated \gls{kf-tp} trajectories. Credible intervals of $2\sigma$ are generated using the arrival time/flown distance at intervals of 500ft for the ensemble of trajectories predicted by the \gls{lwpf}. In the right panels the \gls{tas} estimated by the \gls{lwpf} (blue) and \gls{kf-tp} (red) are compared to observation data and the predicted \gls{tas} from \gls{bada}. 

With the exception of the A320 in climb in Figure~\ref{fig:ens_pred_climb1}, there is often significant misspecification of the \gls{bada} speed profile compared to the observed data. In the case of the climbs in Figure~\ref{fig:ens_pred_climb2} and \ref{fig:ens_pred_climb3} and descent in Figure~\ref{fig:ens_pred_descent3} the discrepancy in \gls{tas} profile appears consistent throughout the trajectory. In other trajectories the discrepancy is more complex: the A320 climb in Figure~\ref{fig:ens_pred_climb1} performs the \gls{cas}-Mach transition around 30,000 ft, which is much lower than the transition altitude in the \gls{bada} model. In Figures~\ref{fig:ens_pred_descent1} and \ref{fig:ens_pred_descent2} the \gls{bada} \gls{tas} profile is expected to be constant above the tropopause (around 36,000 ft), accelerating between the tropopause and \gls{cas}-Mach transition, before decreasing with altitude during the subsequent descent. Although the \gls{bada} \gls{tas} profile can often be significantly misspecified compared to the real trajectory data, the plots demonstrate how the \gls{lwpf} is sufficiently flexible to adjust it's estimate of the state despite the range in aircraft performance observed. The state estimate of both the \gls{kf-tp} and \gls{lwpf} closely tracks the observed Mode S radar, Figure~\ref{fig:ens_pred_descent2} shows an example of the \gls{lwpf} re-initialising due to the particles becoming too concentrated to capture the changing descent mode of the C56X trajectory around 400s. The \gls{kf-tp} tracks the state more closely as the Kalman filter can continuously adjust to new observations, while state estimation in the particle filter is dependent on the particles in the filter being sufficiently diverse to capture the true state of the aircraft.

\begin{figure}
    \centering
    \begin{subfigure}[b]
    {0.49\textwidth}
    \centering
    \includegraphics[width=1.0\linewidth]{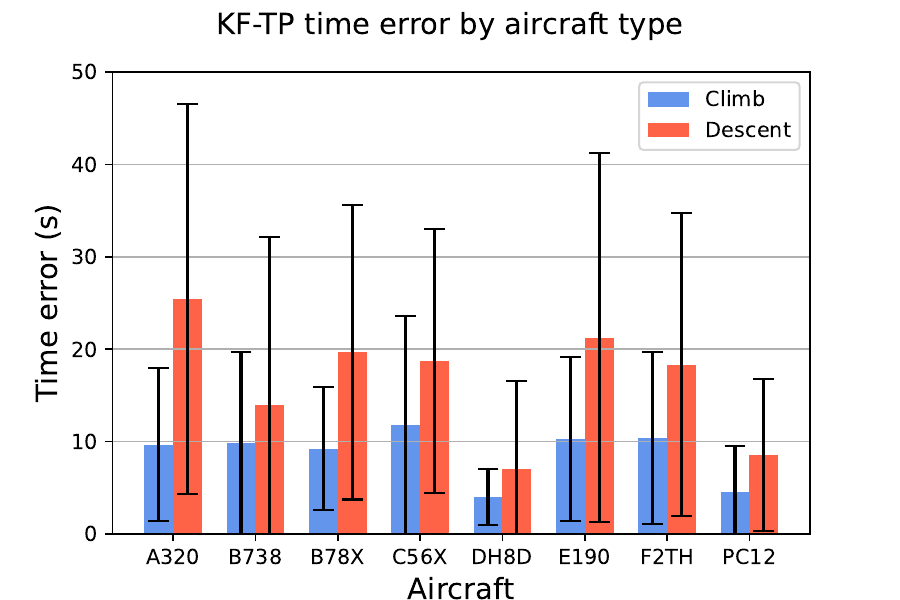}
    \caption{}
    \label{fig:KF_time_actype}
    \end{subfigure}
    \begin{subfigure}[b]
    {0.49\textwidth}
    \centering
    \includegraphics[width=1.0\linewidth]{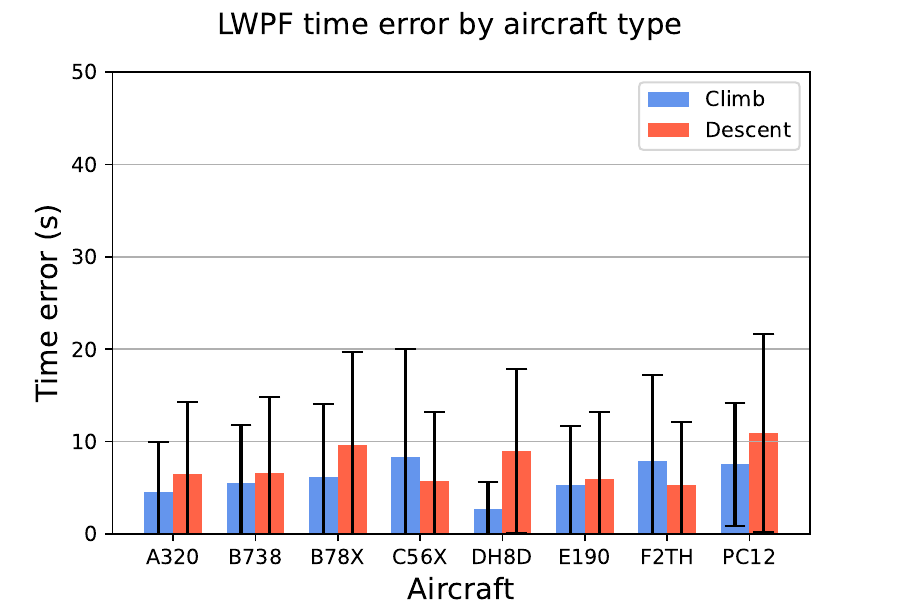}
    \caption{}
    \label{fig:LWPF_time_actype}
    \end{subfigure}
    \begin{subfigure}[b]
    {0.49\textwidth}
    \centering
    \includegraphics[width=1.0\linewidth]{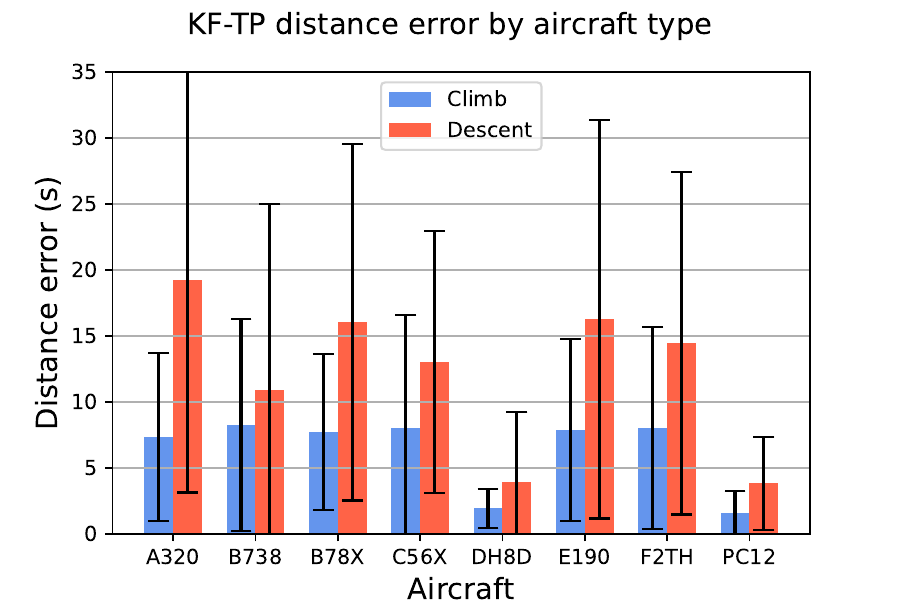}
    \caption{}
    \label{fig:KF_distance_actype}
    \end{subfigure}
    \begin{subfigure}[b]
    {0.49\textwidth}
    \centering
    \includegraphics[width=1.0\linewidth]{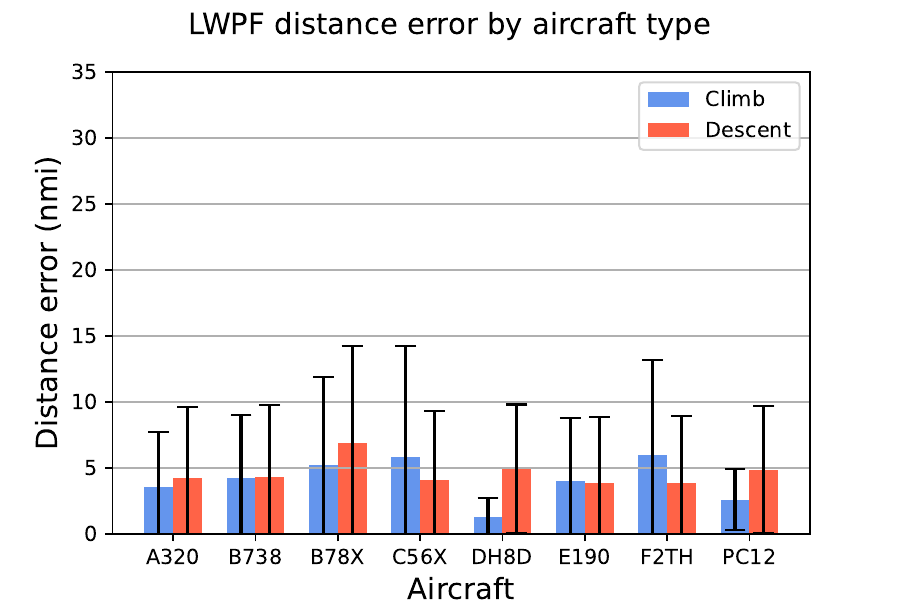}
    \caption{}
    \label{fig:LWPF_distance_actype}
    \end{subfigure}
    \caption{\gls{kf-tp} \gls{mae} compared to \gls{lwpf}, displayed by aircraft type.}
    \label{fig:kf_lwpf_by_actype}
\end{figure}

\begin{table}
\centering
\begin{tabular}{l|cc|cc|cc|cc|cc}
\toprule
Aircraft type & \multicolumn{4}{c|}{MAE time error (s)} & \multicolumn{4}{c}{MAE distance error (nmi)} \\
& \multicolumn{2}{c}{Climb} & \multicolumn{2}{c|}{Descent} & \multicolumn{2}{c}{Climb} & \multicolumn{2}{c}{Descent} \\
& LWPF & KF-TP & LWPF & KF-TP & LWPF & KF-TP & LWPF & KF-TP \\
\midrule
A320 & \cellcolor[RGB]{179,255,179}4.57 & 6.53 & \cellcolor[RGB]{179,255,179}6.44 & 20.36 & \cellcolor[RGB]{179,255,179}3.54 & 5.23 & \cellcolor[RGB]{179,255,179}4.26 & 15.73 \\
B738 & \cellcolor[RGB]{179,255,179}5.57 & 9.85 & \cellcolor[RGB]{179,255,179}6.64 & 13.91 & \cellcolor[RGB]{179,255,179}4.26 & 8.23 & \cellcolor[RGB]{179,255,179}4.30 & 10.92 \\
B78X & 6.27 & \cellcolor[RGB]{179,255,179}5.58 & 9.07 & 14.78 & \cellcolor[RGB]{179,255,179}5.31 & 5.19 & \cellcolor[RGB]{179,255,179}6.71 & 12.13 \\
C56X & \cellcolor[RGB]{179,255,179}8.29 & 9.13 & \cellcolor[RGB]{179,255,179}5.81 & 13.77 & \cellcolor[RGB]{179,255,179}5.78 & 6.45 & \cellcolor[RGB]{179,255,179}4.04 & 9.56 \\
DH8D & \cellcolor[RGB]{179,255,179}2.71 & 3.98 & 8.96 & \cellcolor[RGB]{179,255,179}7.05 & \cellcolor[RGB]{179,255,179}1.31 & 1.93 & 4.92 & \cellcolor[RGB]{179,255,179}3.93 \\
E190 & \cellcolor[RGB]{179,255,179}5.33 & 7.72 & \cellcolor[RGB]{179,255,179}6.00 & 16.67 & \cellcolor[RGB]{179,255,179}4.05 & 6.14 & \cellcolor[RGB]{179,255,179}3.95 & 12.91 \\
F2TH & \cellcolor[RGB]{179,255,179}7.42 & 8.29 & \cellcolor[RGB]{179,255,179}5.10 & 14.41 & \cellcolor[RGB]{179,255,179}5.67 & 6.69 & \cellcolor[RGB]{179,255,179}3.54 & 11.43 \\
PC12 & 7.62 & \cellcolor[RGB]{179,255,179}6.34 & 10.30 & 6.53 & \cellcolor[RGB]{179,255,179}2.64 & 2.25 & 4.64 & \cellcolor[RGB]{179,255,179}2.96 \\
\bottomrule
\end{tabular}
\caption{Comparison of \gls{kf-tp} errors to the \gls{lwpf} by aircraft type.} 
\label{tab:kf_versus_lwpf}
\end{table}

\begin{figure}
    \centering
    \includegraphics[width=0.7\linewidth]{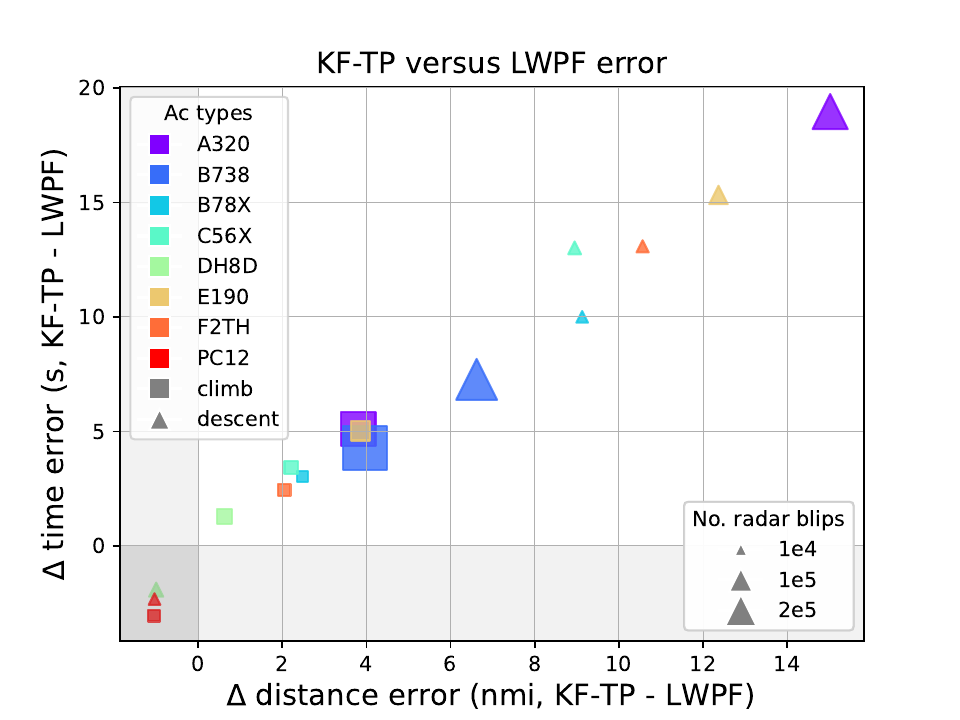}
    \caption{The difference in errors between the \gls{kf-tp} and \gls{lwpf} filtering methods, per aircraft type, with marker size determined by number of radar blips in the test dataset for that aircraft type.}
    \label{fig:kf_versus_lwpf_test_size}
\end{figure}

\begin{figure}
    \centering
    \begin{subfigure}[b]
    {\textwidth}
    \centering
    \includegraphics[width=0.75\linewidth]{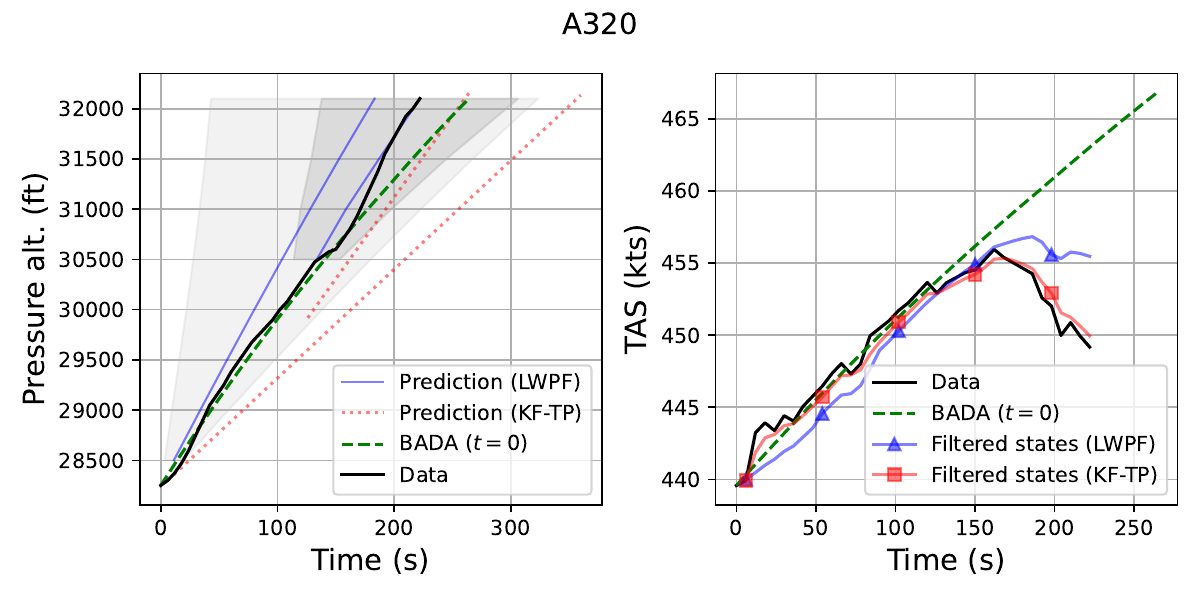}
    \caption{}
    \label{fig:ens_pred_climb1}
    \end{subfigure}
    \begin{subfigure}[b]
    {\textwidth}
    \centering
    \includegraphics[width=0.75\linewidth]{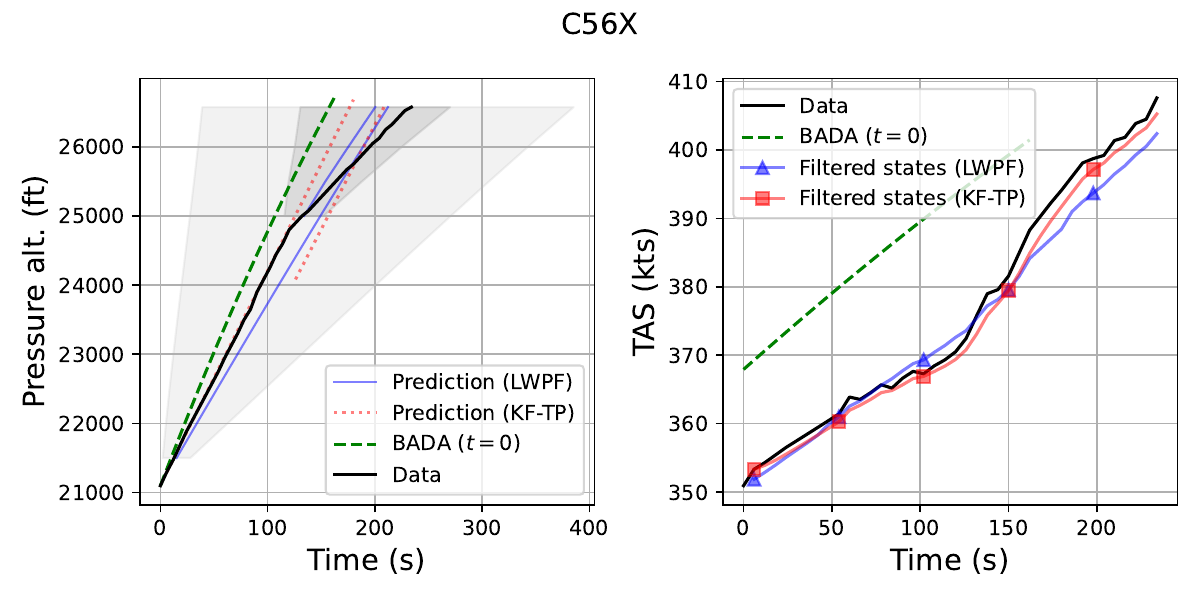}
    \caption{}
    \label{fig:ens_pred_climb2}
    \end{subfigure}
    \begin{subfigure}[b]
    {\textwidth}
    \centering
    \includegraphics[width=0.75\linewidth]{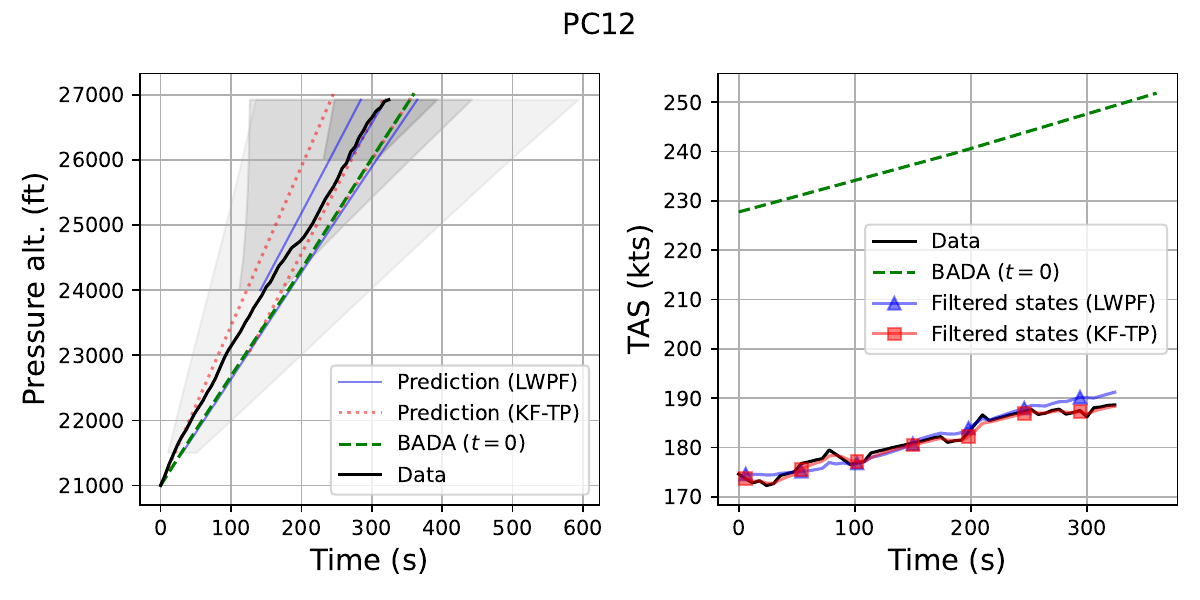}
    \caption{}
    \label{fig:ens_pred_climb3}
    \end{subfigure}
    \caption{Climbing trajectories sampled from the test dataset, compared to \gls{tp}s from the investigated methods. Panels on the left display altitude against time, while those on the right indicate \gls{tas} against time.}
    \label{fig:ens_pred_climb}
\end{figure}

\begin{figure}
    \centering
    \begin{subfigure}[b]
    {\textwidth}
    \centering
    \includegraphics[width=0.75\linewidth]{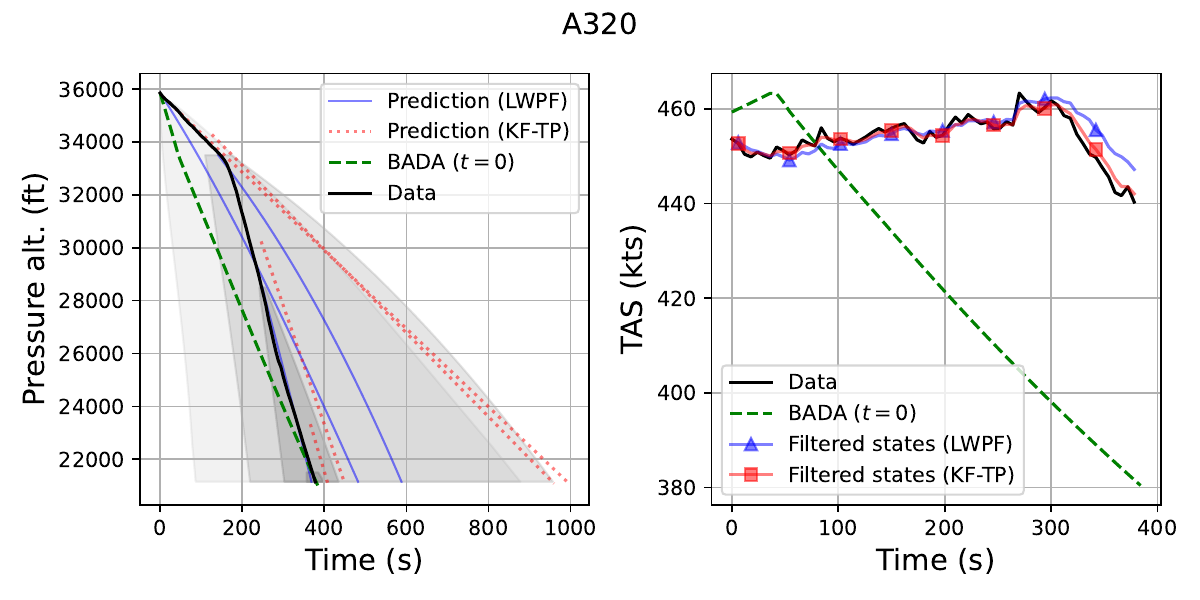}
    \caption{}
    \label{fig:ens_pred_descent1}
    \end{subfigure}
    \begin{subfigure}[b]
    {\textwidth}
    \centering
    \includegraphics[width=0.75\linewidth]{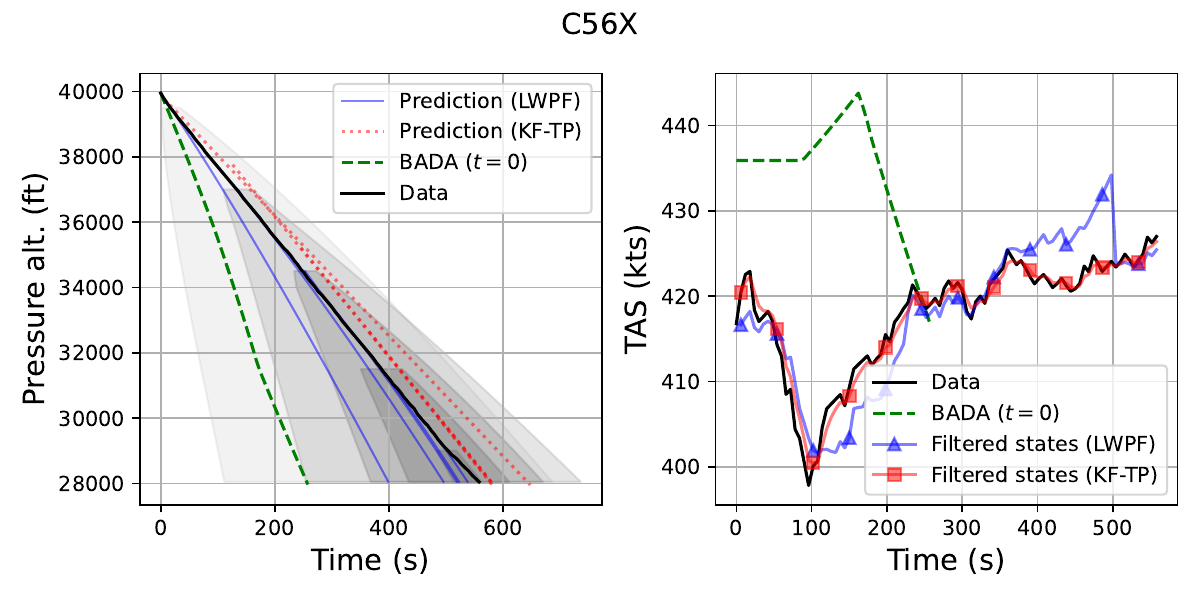}
    \caption{}
    \label{fig:ens_pred_descent2}
    \end{subfigure}
    \begin{subfigure}[b]
    {\textwidth}
    \centering
    \includegraphics[width=0.75\linewidth]{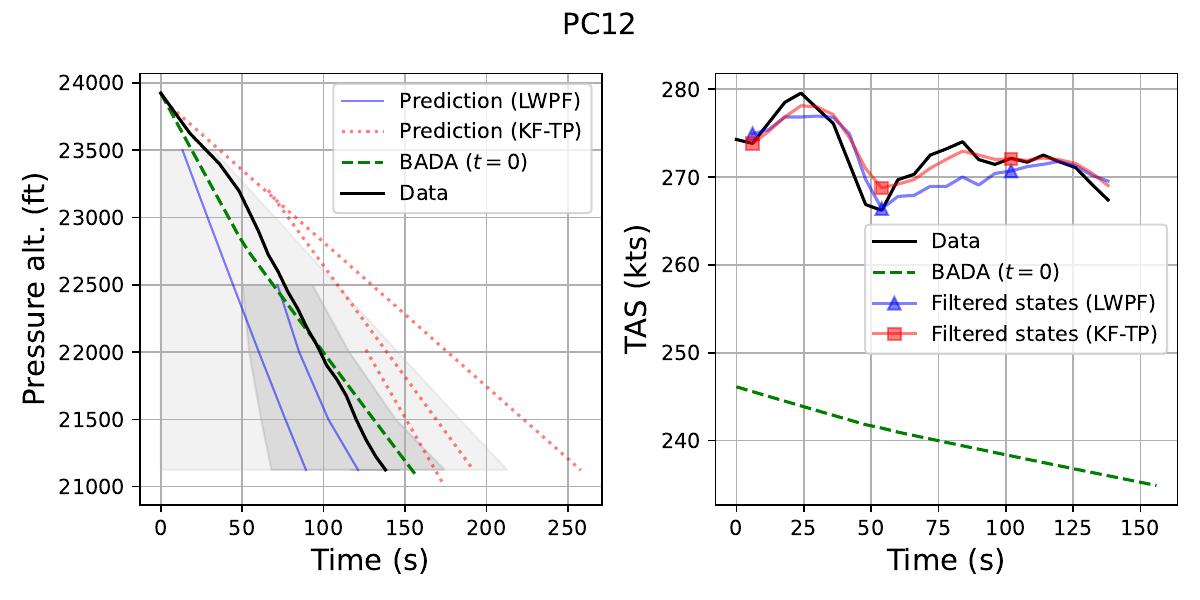}
    \caption{}
    \label{fig:ens_pred_descent3}
    \end{subfigure}
    \caption{Descending trajectories sampled from the test dataset, compared to \gls{tp}s from the investigated methods. Panels on the left display altitude against time, while those on the right indicate \gls{tas} against time.}
    \label{fig:ens_pred_descent}
\end{figure}

\FloatBarrier
\section{Discussion}
A novel method for adaptive \gls{tp} has been presented in this paper. The method is based on a particle filter that uses a set of learned \gls{lssm}s as a prior for aircraft dynamics. The model has demonstrated increased accuracy in climb and descent compared to a set of benchmarks that reflect current state of the art \gls{tp} methods. The proposed method offered time predictions that were 46.3\% and 64.7\% lower than the nearest benchmark for climb and descent respectively. Beyond the presented application to adaptive \gls{tp}, \gls{tp} using \gls{lssm}s offers a fast surrogate for \gls{bada}, an industry standard \gls{tp} method that is based around the numerical solution of a \gls{pde}, by a factor of 5.26 in climb and 8.11 in descent. 

The results presented in this paper are promising, but they also highlight several areas where the method could be improved with future work. Firstly, performance of the method clearly degrades for aircraft types with relatively low numbers of training data. This might be mitigated through a transfer learning approach, in which \gls{lssm}s from other aircraft types with similar dynamics can be pooled. The \gls{lssm}s model \gls{rocd} and \gls{tas} jointly. However, there are instances, such as descending turboprops, where better results can be obtained with a simpler \gls{tp} model. Such cases may well be better handled by an even simpler, approach using polynomials as \gls{tp} surrogates. Finally, although confidence intervals were plotted in Figures~\ref{fig:ens_pred_climb} and Figure~\ref{fig:ens_pred_descent} for illustrative purposes, this paper has been concerned with validating the accuracy of the proposed approach, rather than assessing the calibration of the probabilistic bounds. Further work will develop this aspect of the method, using the frequency with which observed data points lie outside of the predicted bounds as a diagnostic tool that can be used to further improve the model and assist the user to define the contours of its trustworthiness. 

\section*{Appendix}
\subsection{Linearisation of the BADA equations}\label{sec:app_A}

Usually, a two-dimensional system whose dynamics are described by:
\begin{align}
    \dot{\boldsymbol{x}}=
    \begin{bmatrix}
        f(\boldsymbol{x}, \boldsymbol{u}) \\
        g(\boldsymbol{x}, \boldsymbol{u})
    \end{bmatrix},
\end{align}
where $f(\cdot)$ and $g(\cdot)$ are non-linear functions, may be linearized by taking a Taylor series around an operating point, $\boldsymbol{x}_e$. Using the definition of the state in Section~\ref{sec:bada_ssm} the expansion for $f(\cdot)$ is: 

\begin{equation}
    f(h, V_{TAS})\approx f(h_e, V_{e})+\frac{\partial{f}}{\partial h}\Big|_{h_e, V_e}\Delta h+\frac{\partial{f}}{\partial V_{TAS}}\Big|_{h_e, V_e}\Delta V_{TAS}+\frac{\partial f}{\partial u}\Big|_{h_e, V_e}\Delta u,
\end{equation}
where $\boldsymbol{x}_e=[h_e, V_e]^\top$ denotes the operating point. Using an equivalent expression for $g(\cdot)$ the continuous time linear system may be written as:
\begin{equation}
    A=\begin{bmatrix}
        \frac{\partial{f}}{\partial h}\Big|_{h_e, V_e} & \frac{\partial{f}}{\partial V_{TAS}}\Big|_{h_e, V_e}\\
        \frac{\partial{g}}{\partial h}\Big|_{h_e, V_e} & \frac{\partial{g}}{\partial V_{TAS}}\Big|_{h_e, V_e}\\
    \end{bmatrix}
    \; \text{and}\; B=
    \begin{bmatrix}
   \frac{\partial{f}}{\partial u}\Big|_{h_e, V_e}\\ \frac{\partial{g}}{\partial u}\Big|_{h_e, V_e}
    \end{bmatrix}.
    \label{eq:bada_AB}
\end{equation}
$f(h, V_{TAS})$ is the equivalent of \eqref{eq:bada_rocd}. It is natural to treat the aircraft thrust, $T_{HR}$, as the external forcing term. From equations (3.7-1) and (3.7-2) in Nuic et al. \cite{nuic2010user} it is apparent that $T_{HR}$ is a function of $h$ for jet aircraft and $h$ and $V_{TAS}$ for turboprop aircraft such as the DH8D and PC12. In other words, the forcing is a function of the state variables and not time. The \gls{tas} of a climbing aircraft in \gls{bada} in the standard atmosphere can be well approximated as a cubic function of $h$ (see Figure~\ref{fig:tas_fitting}), i.e.:

\begin{equation}
   V_{TAS}=\lambda_1h^3+\lambda_2h^2+\lambda_3h+\lambda_4, 
\end{equation}
with:
\begin{equation}
    g(h, V_{TAS}) = (3\lambda_1 h^2 + 2\lambda_2+\lambda_3)f(h, V_{TAS}),
    \label{eq:g}
\end{equation}
through the chain rule. The forcing in $f(h, V_{TAS})$ rescaled by a quadratic function of $h$ in \eqref{eq:g}. Properly speaking the forcing should be absorbed into $A$ and $B$ set to zero as it is only a function of state parameters. However, it was found that the fit to data was better when a constant forcing with two separate constant forcing parameters was learned as the models had greater flexibility, hence $B\in\Re^2$ in this paper. 

\begin{figure}
    \centering
    \includegraphics[width=0.6\linewidth]{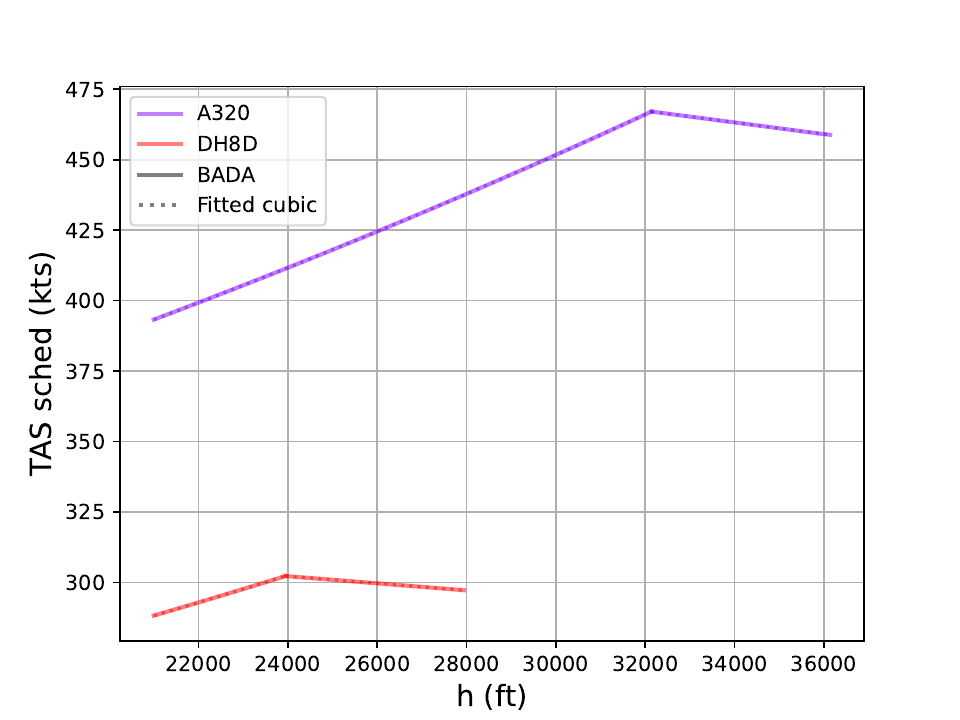}
    \caption{Fitting cubic functions to \gls{tas} as a function of $h$, above and below transition point, for the A320 and DH8D as examples of jet and turboprop aircraft respectively.}
    \label{fig:tas_fitting}
\end{figure}

\FloatBarrier
\subsection{Kalman filter benchmark}\label{sec:app_kalman}
The Kalman filter is a standard approach for state estimation that has been successfully applied in many domains. It is therefore a logical benchmark to measure an algorithm that performs state estimation against. However, the studied application of adaptive \gls{tp} requires both state estimation \emph{and} trajectory prediction. We propose the following benchmark for adaptive \gls{tp} that is based around a Kalman filter, which we denote the \gls{kf-tp}. We define a three-dimensional state that contains the \gls{rocd}, \gls{tas} and altitude of an aircraft. All three of these quantities are returned by Mode S radar returns. A simplistic trajectory prediction is performed by assuming the aircraft will continue to climb or descend at the filtered \gls{rocd} and \gls{tas}. 

Some quantities in the Kalman filter must be set by the user such as the initial state uncertainty matrix, $P$, and the process noise $Q$. In what follows these matrices take the following structure: 

\begin{align}
    P = \alpha_p I_3 \; \text{and} \; Q=\alpha_q I_3,
\end{align}
where $I_3$ is the $3\times3$ identity matrix. The scalars $\alpha_p$ and $\alpha_q$ were scalars that were set by performing a sweep over the validation dataset for different values of these hyperparameters. Precise values for the measurement noise covariance, $R$, are not publicly available. For the purposes of this paper the error covariances were set reasonable estimates of 100 ft/min for the \gls{rocd}, 2.5 knots for \gls{tas}, and 100 ft. Additionally, an external forcing term was included in the state prediction step of the filter of:
\begin{align}
    \Phi_B = \begin{bmatrix}
        0 \\ \alpha_b \\ 0 
    \end{bmatrix},
\end{align}
where $\alpha_b$ was a forcing applied to the aircraft altitude, this is consistent with the scaling of the fitted \gls{lssm}s in Section~\ref{sec:bada_ssm}. Sweeps were performed over the validation set to determine the best choices of hyperparameters for the \gls{kf-tp}. The first swept over values of $\alpha_p$ and $\alpha_q$, while holding the forcing constant, with $\alpha_b =\pm1500\delta t/60$, where $\delta t =6$s refers to the timestep between radar blips. These weights can be interpreted as a constant increase or decrease in altitude per timestep. 

Table~\ref{tab:val_results_kf} tabulates the results of this sweep. The `number of points' column indicates the numbers of blips in the validation set for which a trajectory was successfully generated. A successfully generated trajectory has an absolute \gls{rocd} of greater than or equal to 500 ft/min for all radar blips. This reflects that there is a legal minimum climb or descent rate of 500 ft/min in UK airspace one a clearance for climb or descent has been issued by an air traffic controller. This sweep demonstrated that the performance of the filter was relatively insensitive to the choice of $\alpha_q$. 

A second sweep was performed over $\alpha_b$, with $\alpha_p=\text{1e5}$ and $\alpha_q=\text{1}$, following the results in Table~\ref{tab:val_results_kf}. The results of this sweep are presented in Table~\ref{tab:val_results_kf_B}. Best results across the B738 and DH8D were obtained for $\alpha_b = 500$ft/min in climb, with the results for descent more inconsistent between the two aircraft types. The results in Section~\ref{sec:results} were obtained using $\alpha_b = 500$ft/min in climbs and $\alpha_b = 1500$ft/min in descents. 

\subsection{Sweep over number of particles}\label{sec:app_lwpf}
Table~\ref{tab:particle_sweep} displays how the performance of the \gls{lwpf} varies with number of particles in the filter. As above, the sweep was performed over the B738 and DH8D aircraft types as examples of aircraft with both significantly different performance and different numbers of trajectories in the training and validation datasets. As might be expected, performance improves for the B738 as more particles are included in the filter. However, the reduction in \gls{mae} diminishes with particle size, hence a particle size of 400 was chosen as a compromise between improved \gls{mae} in the validation set and the computational cost associated with generating trajectories for each particle. The trend is more complex for the DH8D, where the \gls{mae} is lower for lower numbers of particles. However, at the same time the failure rate for the trajectory generation is much higher for these particles, for instance 72.6\% of trajectories generated by the filter with 50 particles have an absolute \gls{rocd} of less than 500 ft/min.  Given that there was no clear trend for the DH8D and to ensure that the failure rate was on the order of 5\% or below, a particle size of 400 was used across all aircraft types. 

\begin{landscape}
\begin{table}
\centering
{\footnotesize
\begin{tabular}{llc|cccc|cccc}
\toprule
$P$ scaling & $Q$ scaling & Aircraft type & \multicolumn{4}{c}{Climb} & \multicolumn{4}{c}{Descent} \\
 &  & & Time error (s) & Distance  & Failure & Number of & Time error (s) & Distance  & Failure & Number of \\
  &  & &  & error (nmi) & rate & points &  & error (nmi) &  rate & points \\ 
\midrule
100000 & 0.0001 & B738 & 9.82 & 8.30 & 0.00 & 144156 & 13.13 & 10.21 & 0.01 & 127224 \\
100000 & 0.01 & B738 & 9.75 & 8.07 & 0.00 & 144156 & 12.93 & 9.92 & 0.01 & 127224 \\
100000 & 1.00 & B738 & \cellcolor[RGB]{179,255,179}8.42 & \cellcolor[RGB]{179,255,179}6.82 & 0.00 & 144157 & \cellcolor[RGB]{179,255,179}10.90 & \cellcolor[RGB]{179,255,179}8.17 & 0.01 & 127219 \\
10000 & 0.0001 & B738 & 9.82 & 8.30 & 0.00 & 144156 & 13.14 & 10.22 & 0.01 & 127220 \\
10000 & 0.01 & B738 & 9.76 & 8.07 & 0.00 & 144156 & 12.95 & 9.93 & 0.01 & 127219 \\
10000 & 1.00 & B738 & 8.42 & 6.82 & 0.00 & 144157 & 10.91 & 8.18 & 0.01 & 127216 \\
1000 & 0.0001 & B738 & 9.82 & 8.31 & 0.00 & 144152 & 13.26 & 10.33 & 0.01 & 127177 \\
1000 & 0.01 & B738 & 9.76 & 8.07 & 0.00 & 144152 & 13.07 & 10.04 & 0.01 & 127176 \\
1000 & 1.00 & B738 & \cellcolor[RGB]{179,255,179}8.42 & \cellcolor[RGB]{179,255,179}6.82 & 0.00 & 144153 & 11.00 & 8.26 & 0.01 & 127175 \\
100 & 0.0001 & B738 & 9.88 & 8.35 & 0.00 & 144111 & 14.49 & 11.33 & 0.01 & 126727 \\
100 & 0.01 & B738 & 9.82 & 8.11 & 0.00 & 144112 & 14.28 & 11.02 & 0.01 & 126725 \\
100 & 1.00 & B738 & 8.43 & \cellcolor[RGB]{179,255,179}6.82 & 0.00 & 144109 & 11.87 & 8.97 & 0.01 & 126745 \\
1 & 0.0001 & B738 & 14.72 & 11.62 & 0.01 & 142882 & 42.15 & 30.80 & 0.17 & 106589 \\
1 & 0.01 & B738 & 14.34 & 11.21 & 0.01 & 142947 & 38.83 & 28.07 & 0.16 & 107778 \\
1 & 1.00 & B738 & 9.32 & 7.42 & 0.00 & 143740 & 16.13 & 12.00 & 0.06 & 120026 \\
\midrule
100000 & 0.0001 & DH8D & 5.10 & 2.45 & 0.00 & 8044 & 7.05 & 3.85 & 0.00 & 4708 \\
100000 & 0.01 & DH8D & 5.08 & 2.43 & 0.00 & 8044 & 7.02 & 3.83 & 0.00 & 4708 \\
100000 & 1.00 & DH8D & \cellcolor[RGB]{179,255,179}3.80 & \cellcolor[RGB]{179,255,179}1.83 & 0.00 & 8044 & \cellcolor[RGB]{179,255,179}5.75 & \cellcolor[RGB]{179,255,179}3.17 & 0.00 & 4708 \\
10000 & 0.0001 & DH8D & 5.10 & 2.45 & 0.00 & 8044 & 7.05 & 3.85 & 0.00 & 4708 \\
10000 & 0.01 & DH8D & 5.08 & 2.43 & 0.00 & 8044 & 7.03 & 3.84 & 0.00 & 4708 \\
10000 & 1.00 & DH8D & \cellcolor[RGB]{179,255,179}3.80 & \cellcolor[RGB]{179,255,179}1.83 & 0.00 & 8044 & 5.76 & \cellcolor[RGB]{179,255,179}3.17 & 0.00 & 4708 \\
1000 & 0.0001 & DH8D & 5.11 & 2.45 & 0.00 & 8044 & 7.16 & 3.92 & 0.00 & 4708 \\
1000 & 0.01 & DH8D & 5.09 & 2.44 & 0.00 & 8044 & 7.14 & 3.90 & 0.00 & 4708 \\
1000 & 1.00 & DH8D & \cellcolor[RGB]{179,255,179}3.80 & \cellcolor[RGB]{179,255,179}1.83 & 0.00 & 8044 & 5.86 & 3.23 & 0.00 & 4708 \\
100 & 0.0001 & DH8D & 5.18 & 2.49 & 0.00 & 8044 & 8.18 & 4.50 & 0.00 & 4708 \\
100 & 0.01 & DH8D & 5.16 & 2.47 & 0.00 & 8044 & 8.15 & 4.49 & 0.00 & 4708 \\
100 & 1.00 & DH8D & 3.85 & 1.86 & 0.00 & 8044 & 6.85 & 3.80 & 0.00 & 4708 \\
1 & 0.0001 & DH8D & 7.01 & 3.36 & 0.00 & 8044 & 24.90 & 13.79 & 0.01 & 4659 \\
1 & 0.01 & DH8D & 6.95 & 3.34 & 0.00 & 8044 & 24.62 & 13.64 & 0.01 & 4660 \\
1 & 1.00 & DH8D & 4.43 & 2.14 & 0.00 & 8044 & 15.86 & 8.81 & 0.01 & 4684 \\
\bottomrule
\end{tabular}
}
\caption{Sweep of Kalman filter parameters over the validation datasets for $P$ and $Q$ scalings.}
\label{tab:val_results_kf}
\end{table}
\end{landscape}

\begin{landscape}
\begin{table}
\centering
{\footnotesize
\begin{tabular}{lc|cccc|cccc}
\toprule
$B$ scaling & Aircraft type & \multicolumn{4}{c}{Climb} & \multicolumn{4}{c}{Descent} \\
   & & Time error (s) & Distance  & Failure & Number of & Time error (s) & Distance  & Failure & Number of \\
    & &  & error (nmi) & rate & points &  & error (nmi) &  rate & points \\ 
\midrule
2000 & B738 & 9.45 & 7.56 & 0.00 & 144157 & \cellcolor[RGB]{179,255,179}10.65 & \cellcolor[RGB]{179,255,179}7.84 & 0.01 & 127239 \\
1500 & B738 & 8.42 & 6.82 & 0.00 & 144157 & 10.90 & 8.17 & 0.01 & 127219 \\
1000 & B738 & 7.67 & 6.27 & 0.00 & 144152 & 11.51 & 8.72 & 0.01 & 127200 \\
500 & B738 & \cellcolor[RGB]{179,255,179}7.49 & \cellcolor[RGB]{179,255,179}6.14 & 0.00 & 144148 & 12.50 & 9.47 & 0.01 & 127165 \\
0 & B738 & 7.71 & 6.29 & 0.00 & 144144 & 13.70 & 10.33 & 0.01 & 127142 \\
\midrule
2000 & DH8D & 5.17 & 2.47 & 0.00 & 8044 & 5.97 & 3.26 & 0.00 & 4708 \\
1500 & DH8D & 3.80 & 1.83 & 0.00 & 8044 & \cellcolor[RGB]{179,255,179}5.75 & \cellcolor[RGB]{179,255,179}3.17 & 0.00 & 4708 \\
1000 & DH8D & 2.38 & 1.17 & 0.00 & 8044 & 6.34 & 3.51 & 0.00 & 4708 \\
500 & DH8D & \cellcolor[RGB]{179,255,179}2.34 & \cellcolor[RGB]{179,255,179}1.14 & 0.00 & 8044 & 7.93 & 4.41 & 0.00 & 4708 \\
0 & DH8D & 3.61 & 1.73 & 0.00 & 8044 & 10.16 & 5.63 & 0.00 & 4708 \\
\bottomrule
\end{tabular}
\caption{Sweep of Kalman filter parameters over the validation datasets for the $B$ scaling.}
\label{tab:val_results_kf_B}}
\end{table}
\end{landscape}

\begin{landscape}
\begin{table}
\begin{tabular}{lc|cccc|ccccc}
\toprule
Number of & Aircraft type & \multicolumn{4}{c}{Climb} & \multicolumn{4}{c}{Descent} \\
 particles &   & Time error (s) & Distance  & Failure & Number of & Time error (s) & Distance  & Failure & Number of \\
 & &  & error (nmi) & rate & points &  & error (nmi) &  rate & points \\
\midrule
50 & B738 & 6.43 & 5.09 & 0.01 & 143437 & 7.90 & 5.08 & 0.00 & 127611 \\
100 & B738 & 5.90 & 4.62 & 0.00 & 144171 & 7.41 & 4.82 & 0.00 & 128068 \\
200 & B738 & 5.59 & 4.33 & 0.00 & 144193 & 6.94 & 4.53 & 0.00 & 128136 \\
400 & B738 & 5.44 & 4.18 & 0.00 & 144281 & 6.67 & 4.33 & 0.00 & 128207 \\
800 & B738 & 5.38 & 4.11 & 0.00 & 144292 & 6.53 & 4.26 & 0.00 & 128231 \\
1600 & B738 & \cellcolor[RGB]{179,255,179}5.33 & \cellcolor[RGB]{179,255,179}4.06 & 0.00 & 144297 & 6.41 & 4.17 & 0.00 & 128167 \\
3200 & B738 & 5.34 & \cellcolor[RGB]{179,255,179}4.06 & 0.00 & 144297 & \cellcolor[RGB]{179,255,179}6.35 & \cellcolor[RGB]{179,255,179}4.15 & 0.00 & 128243 \\
\midrule
50 & DH8D & 3.74 & 1.77 & 0.00 & 8039 & \cellcolor[RGB]{255,255,0}4.41 & \cellcolor[RGB]{255,255,0}2.41 & 0.73 & 1288 \\
100 & DH8D & \cellcolor[RGB]{179,255,179}3.61 & \cellcolor[RGB]{179,255,179}1.71 & 0.00 & 8044 & 6.13 & 3.37 & 0.44 & 2634 \\
200 & DH8D & 3.68 & 1.74 & 0.00 & 8044 & 7.61 & 4.17 & 0.22 & 3657 \\
400 & DH8D & 3.63 & 1.72 & 0.00 & 8044 & 9.21 & 5.07 & 0.05 & 4470 \\
800 & DH8D & 3.62 & \cellcolor[RGB]{179,255,179}1.71 & 0.00 & 8044 & 9.52 & 5.23 & 0.01 & 4650 \\
1600 & DH8D & 3.65 & 1.73 & 0.00 & 8044 & 9.52 & 5.24 & 0.00 & 4700 \\
\bottomrule
\end{tabular}
\caption{Sweep over particle sizes in the \gls{lwpf} on the validation dataset}
\label{tab:particle_sweep}
\end{table}

\end{landscape}
\FloatBarrier
\section*{Acknowledgments}
The authors would like to thank Dr. Lawrence Bull, Research Fellow at the University of Glasgow, for providing useful insight on particle filtering. 

\section*{Funding Sources}
The work described in this paper is primarily funded by the grant “EP/V056522/1: Advancing Probabilistic Machine Learning to Deliver Safer, More Efficient and Predictable Air Traffic Control” (aka Project Bluebird), an EPSRC Prosperity Partnership between NATS, The Alan Turing Institute, the University of Exeter, and the University of Cambridge. 


\bibliography{data_assim}

@article{nuic2010user,
  title={User manual for the Base of Aircraft Data (BADA) revision 3.10},
  author={Nuic, Angela},
  journal={Atmosphere},
  volume={2010},
  pages={001},
  year={2010},
  url={{http://maartenuijtdehaag.com/bada310-user-manual.pdf}}
}

@phdthesis{lymperopoulos2010sequential,
  title={Sequential Monte Carlo methods in air traffic management},
  author={Lymperopoulos, Ioannis},
  year={2010},
  school={ETH Zurich},
  url = {{https://www.research-collection.ethz.ch/bitstream/handle/20.500.11850/152184/eth-1702-02.pdf}}
}

@article{liu2011,
author = {Liu, Weiyi and Hwang, Inseok},
title = {Probabilistic Trajectory Prediction and Conflict Detection for Air Traffic Control},
journal = {Journal of Guidance, Control, and Dynamics},
volume = {34},
number = {6},
pages = {1779-1789},
year = {2011},
url = {{https://doi.org/10.2514/1.53645}}
}

@article{zhang2015,
author = {Zhang, Junfeng and Wu, X.-G and Wang, F.},
year = {2015},
month = {04},
pages = {180-184},
title = {Aircraft trajectory prediction based on modified interacting multiple model algorithm},
volume = {32},
journal = {Journal of Donghua University (English Edition)},
url = {https://www.researchgate.net/publication/282919671_Aircraft_trajectory_prediction_based_on_modified_interacting_multiple_model_algorithm}
}

@ARTICLE{mazor1998,
  author={Mazor, E. and Averbuch, A. and Bar-Shalom, Y. and Dayan, J.},
  journal={IEEE Transactions on Aerospace and Electronic Systems}, 
  title={Interacting multiple model methods in target tracking: a survey}, 
  year={1998},
  volume={34},
  number={1},
  pages={103-123},
  url = {{https://ieeexplore.ieee.org/document/640267}}
}

@article{Bayes+weather,
  title   = {Data-driven trajectory prediction with weather uncertainties: A {B}ayesian deep learning approach},
  journal = {Transportation Research Part C: Emerging Technologies},
  volume  = {130},
  pages   = {103326},
  year    = {2021},
  issn    = {0968-090X},
  url    = {https://doi.org/10.1016/j.trc.2021.103326},
  author  = {Yutian Pang and Xinyu Zhao and Hao Yan and Yongming Liu}
}

@article{bastas2020data,
  title={Data driven aircraft trajectory prediction with deep imitation learning},
  author={Bastas, Alevizos and Kravaris, Theocharis and Vouros, George A},
  journal={arXiv},
  year={2020},
  url = {https://arxiv.org/abs/2005.07960}
}

@ARTICLE{silvestre,
  author={Silvestre, Jorge and Mielgo, Paula and Bregon, Anibal and Martínez-Prieto, Miguel A. and Álvarez-Esteban, Pedro C.},
  journal={IEEE Access}, 
  title={Multi-Route Aircraft Trajectory Prediction Using Temporal Fusion Transformers}, 
  year={2024},
  volume={12},
  pages={174094-174106},
  url={https://ieeexplore.ieee.org/abstract/document/10577632}, 
 }

@inproceedings{pred_analytics,
  author    = {Ayhan, Samet and Samet, Hanan},
  title     = {Aircraft Trajectory Prediction Made Easy with Predictive Analytics},
  year      = {2016},
  isbn      = {9781450342322},
  publisher = {Association for Computing Machinery},
  address   = {New York, NY, USA},
  url={https://doi.org/10.1145/2939672.293969},
  booktitle = {Proceedings of the 22nd ACM SIGKDD International Conference on Knowledge Discovery and Data Mining},
  pages     = {21--30},
  numpages  = {10},
  location  = {San Francisco, California, USA},
  series    = {KDD '16}
}

@article{WU2022103554,
title = {Long-term 4D trajectory prediction using generative adversarial networks},
journal = {Transportation Research Part C: Emerging Technologies},
volume = {136},
pages = {103554},
year = {2022},
issn = {0968-090X},
url = {https://www.sciencedirect.com/science/article/pii/S0968090X22000031},
author = {Xiping Wu and Hongyu Yang and Hu Chen and Qinzhi Hu and Haoliang Hu},
}

@article{N-Incept,
  author  = {Chen, Yuchao and Sun, Jinlong and Lin, Yun and Gui, Guan and Sari, Hikmet},
  journal = {IEEE Transactions on Intelligent Transportation Systems},
  title   = {Hybrid N-Inception-LSTM-Based Aircraft Coordinate Prediction Method for Secure Air Traffic},
  year    = {2022},
  volume  = {23},
  number  = {3},
  pages   = {2773--2783},
  url     = {https://doi.org/10.1109/TITS.2021.3095129}
}

@ARTICLE{gmm,
  author={Barratt, Shane T. and Kochenderfer, Mykel J. and Boyd, Stephen P.},
  journal={IEEE Transactions on Intelligent Transportation Systems}, 
  title={Learning Probabilistic Trajectory Models of Aircraft in Terminal Airspace From Position Data}, 
  year={2019},
  volume={20},
  number={9},
  pages={3536-3545},
  url = {https://ieeexplore.ieee.org/document/8551278}}

@article{Pepper_DataBADA,
author = {Pepper, Nick and Thomas, Marc},
title = {Learning Generative Models for Climbing Aircraft from Radar Data},
journal = {Journal of Aerospace Information Systems},
volume = {21},
number = {6},
pages = {474-481},
year = {2024},
url = {https://doi.org/10.2514/1.I011359},
}

@misc{hodgkin2025probabilisticsimulationaircraftdescent,
      title={Probabilistic Simulation of Aircraft Descent via a Physics-Informed Machine Learning Approach}, 
      author={Amy Hodgkin and Nick Pepper and Marc Thomas},
      year={2025},
      eprint={2504.02529},
      archivePrefix={arXiv},
      primaryClass={eess.SY},
      url={https://arxiv.org/abs/2504.02529}, 
}

@article{hsu1981evaluation,
  title={The evaluation of aircraft collision probabilities at intersecting air routes},
  author={Hsu, DA},
  journal={The Journal of Navigation},
  volume={34},
  number={1},
  pages={78--102},
  year={1981},
  publisher={Cambridge University Press},
  url = {https://doi.org/10.1017/S0373463300024279},
}

@article{Anderson_Lin_1996, 
title={A Collision Risk Model for a Crossin Track Separation Methodology},
volume={49}, 
url = {https://doi.org/10.1017/S0373463300013576}, 
number={3}, 
journal={Journal of Navigation}, 
author={Anderson, D. and Lin, X. G.}, 
year={1996}, 
pages={337–349}}

@article{WANG2025109918,
title = {Developing an aircraft takeoff mass estimation model based on the hybrid KMI-DNN-BI model using quick access recorder (QAR) data},
journal = {Aerospace Science and Technology},
volume = {158},
pages = {109918},
year = {2025},
issn = {1270-9638},
url = {https://www.sciencedirect.com/science/article/pii/S1270963824010472},
author = {Bing Wang and Runyuan Zou and Jianfeng Mao and Cheng-Lung Wu and Dabin Xue},
}

@inproceedings{sun2016modeling,
  title={Modeling and inferring aircraft takeoff mass from runway ADS-B data},
  author={Sun, Junzi and Ellerbroek, Joost and Hoekstra, Jacco},
  booktitle={7th International Conference on Research in Air Transportation},
  year={2016},
    pages = {},
  url = {https://pure.tudelft.nl/ws/files/8929936/Sun_Ellerbroek_Hoekstra_2016_Modeling_and_Inferring_Aircraft_Takeoff_Mass_from_Runway_ADS_B_Data.pdf}
}

@article{xiang_and_chen,
author = {Xiang, Jun and Chen, Jun},
title = {Data-Driven Probabilistic Trajectory Learning with High Temporal Resolution in Terminal Airspace},
journal = {Journal of Aerospace Information Systems},
volume = {0},
number = {0},
pages = {1-11},
year = {2025},
URL = {https://doi.org/10.2514/1.I011545},
}

@inproceedings{lymperopoulos2008adaptive,
  title={Adaptive aircraft trajectory prediction using particle filters},
  author={Lymperopoulos, Ioannis and Lygeros, John},
  booktitle={AIAA Guidance, Navigation and Control Conference and Exhibit},
  pages={7387},
  year={2008},
  url = {https://doi.org/10.2514/6.2008-7387}
}

@inproceedings{madyastha2011extended,
  title={Extended Kalman filter vs. error state Kalman filter for aircraft attitude estimation},
  author={Madyastha, Venkatesh and Ravindra, Vishal and Mallikarjunan, Srinath and Goyal, Anup},
  booktitle={AIAA Guidance, Navigation, and Control Conference},
  pages={6615},
  year={2011},
  url = {https://doi.org/10.2514/6.2011-6615}
}

@article{maeder,
author = {Maeder, Urban and Morari, Manfred and Baumgartner, Thomas Ivar},
title = {Trajectory Prediction for Light Aircraft},
journal = {Journal of Guidance, Control, and Dynamics},
volume = {34},
number = {4},
pages = {1112-1119},
year = {2011},
URL = {https://doi.org/10.2514/1.52124},
}

@inbook{schultz,
author = {Charles Schultz and David Thipphavong and Heinz Erzberger},
title = {Adaptive Trajectory Prediction Algorithm for Climbing Flights},
booktitle = {AIAA Guidance, Navigation, and Control Conference},
URL = {https://arc.aiaa.org/doi/abs/10.2514/6.2012-4931},
year = {2012}
}

@incollection{liu2001combined,
  title={Combined parameter and state estimation in simulation-based filtering},
  author={Liu, Jane and West, Mike},
  booktitle={Sequential Monte Carlo methods in practice},
  pages={197--223},
  year={2001},
  publisher={Springer},
  url = {https://doi.org/10.1007/978-1-4757-3437-9_10}
}

@article{nemeth2013sequential,
  title={Sequential Monte Carlo methods for state and parameter estimation in abruptly changing environments},
  author={Nemeth, Christopher and Fearnhead, Paul and Mihaylova, Lyudmila},
  journal={IEEE Transactions on Signal Processing},
  volume={62},
  number={5},
  pages={1245--1255},
  year={2013},
  publisher={IEEE},
  url ={https://ieeexplore.ieee.org/abstract/document/6692890},
}

@article{nelder1965simplex,
  title={A simplex method for function minimization},
  author={Nelder, John A and Mead, Roger},
  journal={The computer journal},
  volume={7},
  number={4},
  pages={308--313},
  year={1965},
  publisher={The British Computer Society},
  url = {https://doi.org/10.1093/comjnl/7.4.308},
}

@inproceedings{DTpaper,
    author = {Nick Pepper and Adam Keane and Amy Hodgkin and Dewi Gould, Edward Henderson and Lynge Lauritsen and Christos Vlahos and George {De Ath} and Richard Everson and Richard Cannon and Alvaro Sierra Castro and John Korna and Ben Carvell and Marc Thomas},
    title = {A Probabilistic Digital Twin of UK En Route Airspace for Training and Evaluating AI Agents for Air Traffic Control},
    booktitle = {AIAA SCITECH 2026 Forum},
    chapter = {},
    pages = {},
    doi = {},
    URL = {},
    eprint = {},
    year = {2026},
}

\end{document}